\documentclass[authoryear]{elsarticle}

\usepackage{geometry}                
\usepackage{graphicx}
\usepackage{amssymb}
\usepackage{epstopdf}
\usepackage{fullpage}
\usepackage{array}
\usepackage{float}
\usepackage{latexsym}
\usepackage{diagbox}
\usepackage{amsfonts}
\usepackage{cool}
\usepackage{dcolumn}
\usepackage{textcomp} 
\usepackage{xcolor}
\usepackage{scrextend}
\usepackage{mathrsfs}
\usepackage{multirow}
\usepackage{amsmath}
\usepackage{enumerate}
\usepackage{cancel}
\usepackage{parskip}
\usepackage{wasysym}
\usepackage{longtable}
\usepackage{lipsum}
\usepackage[authoryear]{natbib}
\usepackage[normalem]{ulem}
\usepackage{changepage} 
\usepackage{tabulary}
\usepackage{rotating}
\usepackage{hyperref}
\hypersetup{colorlinks=true, breaklinks=true}
\usepackage[xindy,toc,section=section,acronym, numberedsection]{glossaries}
\makenoidxglossaries
\setglossarystyle{list}

\newacronym{chime}{CHIME}{Canadian Hydrogen Intensity Mapping Experiment}
\newacronym{hirax}{HIRAX}{Hydrogen Intensity and Real Time Analysis Experiment}
\newacronym{frb}{FRB}{Fast Radio Burst}
\newacronym{dm}{DM}{dispersion measure}
\newacronym{rm}{RM}{rotation measure}
\newacronym{flat}{Fermi LAT}{Fermi Large Area Telescope}
\newacronym{utmost}{UTMOST}{Molonglo Observatory Synthesis Telescope}
\newacronym{ska}{SKA}{Square Kilometre Array}
\newacronym{askap}{ASKAP}{Australian Square Kilometre Array Pathfinder}
\newacronym{apertif}{Apertif}{Apertif Radio Transient System}
\newacronym{wsrt}{WSRT}{Westerbork Synthesis Radio Telescope}
\newacronym{fast}{FAST}{Five-hundred-meter Aperture Spherical radio Telescope}
\newacronym{lofar}{LOFAR}{Low-Frequency Array}
\newacronym{gbm}{Fermi GBM}{Fermi Gamma-ray Burst Monitor}
\newacronym{bat}{Swift/BAT}{Swift Burst Alert Telescope}
\newacronym{igm}{IGM}{intergalactic medium}
\newacronym{smbh}{SMBH}{supermassive black hole}
\newacronym{agn}{AGN}{active galactic nuclei}
\newacronym{grb}{GRB}{gamma ray burst}
\newacronym{sgrb}{sGRB}{short gamma ray burst}
\newacronym{lgrb}{LGRB}{long gamma ray burst}
\newacronym{ns}{NS}{neutron star}
\newacronym[longplural={magnetar wind nebulae},shortplural={MWNe}]{mwn}{MWN}{magnetar wind nebula}
\newacronym{bh}{BH}{black hole}
\newacronym{kbh}{KBH}{Kerr black hole}
\newacronym{knbh}{KNBH}{Kerr-Newman black hole}
\newacronym{vlbi}{VLBI}{very long baseline interferometry}
\newacronym{evn}{EVN}{European VLBI Network}
\newacronym{vla}{VLA}{Very Large Array}
\newacronym{snr}{SNR}{supernova remnant}
\newacronym{em}{EM}{electromagnetic}
\newacronym{dsr}{DSR}{Dicke's superradiance}
\newacronym{wd}{WD}{white dwarf}
\newacronym{sqm}{SQM}{strange quark matter}
\newacronym{mwd}{MWD}{magnetic white dwarf}
\newacronym[longplural={pulsar wind nebulae},shortplural={PWNe}]{pwn}{PWN}{pulsar wind nebula}
\newacronym[longplural={supernovae},shortplural={SNe}]{sn}{SN}{supernova}
\newacronym{pbh}{PBH}{primordial black hole}
\newacronym{sgr}{SGR}{soft gamma repeater}
\newacronym{blast}{BLAST}{black hole laser powered by axion superradiant instabilities}
\newacronym{aqn}{AQN}{axion quark nugget}
\newacronym{lsd}{LSD}{large superconducting dipole}
\newacronym{ism}{ISM}{interstellar medium}
\newacronym{scs}{SCS}{superconducting cosmic string}
\newacronym{cmb}{CMB} {cosmic microwave background}
\newacronym{gw}{GW}{gravitational wave}
\newacronym{ss}{SS}{strange quark star}
\newacronym[longplural={superluminous supernovae},shortplural={SLSNe}]{slsn}{SLSN I}{superluminous supernova}
\newacronym{wh}{WH}{white hole}
\begin{document}
\begin{frontmatter}
	
\title{A Living Theory Catalogue for Fast Radio Bursts}

\author[add1]{E. Platts\corref{cor1}}
\ead{pltemm002@myuct.ac.za}
\author[add1]{A. Weltman}
\ead{amanda.weltman@uct.ac.za}
\author[add2,add3]{A. Walters}
\ead{waltersa@ukzn.ac.za}
\author[add4]{S. P. Tendulkar}
\ead{shriharsh@physics.mcgill.ca}
\author[add1]{J.E.B. Gordin}
\ead{grdjak001@myuct.ac.za}
\author[add1]{S. Kandhai}
\ead{kndsul001@myuct.ac.za}
\cortext[cor1]{Corresponding author}

\address[add1]{High Energy Physics, Cosmology \& Astrophysics Theory (HEPCAT) group, Department of Mathematics and Applied Mathematics, University of Cape Town, Rondebosch, 7700, South Africa}
\address[add2]{Astrophysics \& Cosmology Research Unit, School of Chemistry and Physics, University of KwaZulu-Natal, Durban, 4000, South Africa}
\address[add3]{NAOC-UKZN Computational Astrophysics Centre (NUCAC), University of KwaZulu-Natal, Durban, 4000, South Africa}
\address[add4]{Department of Physics \& McGill Space Institute, McGill University, 3600 University Street, Montreal QC, H3A 2T8, Canada}

\date{\today}

\begin{abstract}
At present, we have almost as many theories to explain Fast Radio Bursts as we have Fast Radio Bursts observed. This landscape will be changing rapidly with CHIME/FRB, recently commissioned in Canada, and HIRAX, under construction in South Africa. This is an opportune time to review existing theories and their observational consequences, allowing us to efficiently curtail viable astrophysical models as more data becomes available. In this article we provide a currently up to date catalogue of the numerous and varied theories proposed for Fast Radio Bursts so far. We also launched an \href{https://frbtheorycat.org/index.php/Main_Page}{online evolving repository} for the use and benefit of the community to dynamically update our theoretical knowledge and discuss constraints and uses of Fast Radio Bursts. 
\end{abstract}


\end{frontmatter}

\newpage
\tableofcontents
\clearpage

\section{Introduction}
A little over a decade after their discovery \citep{Lorimer:2007qn}, \acrfullpl{frb} remain an enigmatic class of radio transients. \acrshortpl{frb} are characterized by one or multiple very bright ($\sim\mathrm{Jy}$) and very brief ($\sim\mathrm{ms}$) bursts of radio photons, and have been detected at frequencies ranging between $400~ \mathrm{MHz} - 8~\mathrm{GHz}$ by a number of ground-based radio telescopes. Importantly, the arrival time of the frequency components is dispersed, precisely going as $\Delta t \sim \nu^{-2}$, which is consistent with the propagation of a radio wave through cold plasma \citep{Lorimer:2007qn,Dennison:2014vea,Caleb:2015uuk}. While some early speculation considered \acrshortpl{frb} to be of galactic origin \citep{Keane:2012yh, Burke-Spolaor:2014rqa, Bannister:2014xla, Maoz:2015xpa}, current consensus is that their excessively large \acrfull{dm} (see Section \ref{dm}), high galactic latitude, apparent isotropy over the sky \citep{Champion:2015pmj}, and lack of $\mathrm{H_{II}}$ regions or other sources of excess \acrshort{dm} indicate an extragalactic \citep{Keane:2012yh,2015RAA....15.1629X,Xu:2015wza, Cordes:2015fua, Connor:2015era} or cosmological \citep{Dolag:2014bca,Kulkarni:2014vea,Katz:2015mpa,Caleb:2015uuk,Vedantham:2016tjy,Cao:2017ucx,Niino:2018xze} origin. This consensus is supported by the identification of the host galaxy of FRB 121102 at redshift z=0.1932 \citep{Chatterjee:2017dqg, Tendulkar:2017vuq}. 

One of the central challenges facing theoretical model builders is finding a physical mechanism with which one can explain the vast amount of energy radiated over such short timescales. If one assumes isotropic emission, the extreme brightness indicates that some beamed, coherent emission process is required \citep{Thornton:2013iua}, and the brevity of the signals suggests the source is extremely compact \citep{Thornton:2013iua}. Compounding the model building challenges, the characteristic properties of \acrshortpl{frb} appear to be heterogeneous. Where measurements have been possible, \acrshortpl{frb} have been observed to have circular \citep{Ravi:2014mma, Petroff:2014taa,Caleb:2018oyn} and/or linear \citep{Petroff:2015bua,Masui:2015kmb,Ravi:2016kfj,Michilli:2018zec,Gajjar:2018bth} polarizations, as well as some that seem unpolarized (although this may be due to extremely high Faraday rotation \citep{Michilli:2018zec}). The pulse profiles of \acrshortpl{frb} also differ: two have double or triple peaks \citep{Champion:2015pmj,Farah:2018buz}, while the rest have only single peaks. Many \acrshortpl{frb} have now shown complex microstructure and features at timescales of 10s of microseconds \citep{Farah:2018buz} (and Hessels \textit{et al.}, \emph{in preparation}). Even more baffling is that only two \acrshortpl{frb}, FRB 121102 and FRB 180814.J0422+73 have been observed to repeat \citep{Spitler:2016dmz,Scholz:2017kwy, Gajjar:2018bth, Spitler:2018sjn, CHIMEFRB:2019b}, with modulating pulse shapes and no apparent periodicity (see Section \ref{curiouscase}). Some \acrshortpl{frb} have been monitored for up to hundreds of hours with no indication of repetition \citep{Lorimer:2007qn,Petroff:2014taa,Ravi:2016kfj,Petroff:2017pff,Bhandari:2017qrj,Shannon:2018frb}, and while this may imply there are two different classes of \acrshort{frb} (repeaters and non-repeaters) \citep{Palaniswamy:2017aze, Michilli:2018zec}, there may just be a large range in repetition rates \citep{Caleb:2019szc}. Further, the observed repetition or lack thereof can be strongly affected by interstellar scintillations \citep{Cordes:1998} or plasma lensing \citep{Cordes:2017eug,Main:2018kfc,Spitler:2018sjn}. FRB 121102 was found to have a \acrfull{rm} 400 times larger than any other known \acrshort{frb} \citep{Michilli:2018zec,Gajjar:2018bth}, but previous \acrshortpl{frb} may have had high \acrshortpl{rm} that were simply not detectable \citep{Petroff:2014taa, Keane:2016yyk}. There is currently no consensus on the matter.

There have only been 54\footnote{Correct on the date of submission, almost doubled in the week of submission \citep{Shannon:2018frb}!} \acrshort{frb} detections subsequent to the Lorimer burst in 2007\footnote{See the online FRB catalogue \citep{Petroff:2016tcr}, found at \url{www.frbcat.org}, for an up-to-date list.}, but the non-detection of \acrshortpl{frb} can provide some insight. For example, one can constrain event rates \citep{Siemion:2011fa,Wayth:2012gm,Trott:2013uwk,Trott:2013gb, Tingay:2015wdv,Karastergiou:2015zsa, Burke-Spolaor:2016sie,Rowlinson:2016zjw, Amiri:2017qtx, Surnis:2017roy}, spectral indices \citep{Tingay:2015wdv,Karastergiou:2015zsa,Burke-Spolaor:2016sie}, and surface densities of transients \citep{Rowlinson:2016zjw}. Arguably, the most illuminating detection to date is (the repeating) FRB 121102 \citep{Spitler:2016dmz}. This is the only event to have been successfully associated with a host galaxy: a low-metallicity star-forming dwarf galaxy at a redshift of $z=0.1932$ \citep{Chatterjee:2017dqg, Tendulkar:2017vuq, Bassa:2017tke}. This confirms that (at least some) \acrshortpl{frb} propagate over extragalactic or cosmological distances.

Apart from being interesting in and of themselves, \acrshortpl{frb} could prove to be very useful cosmological and astrophysical probes. There are already a number of proposals \acrshortpl{frb} to study fundamental cosmological parameters \citep{Zhou:2014yta, Yang:2016zbm, Walters:2017afr, Yu:2017beg,Li:2017mek, Zitrin:2018let, Wei:2018cgd}, extragalactic magnetic fields \citep{Vazza:2018fbb}, properties of the \acrfull{igm} \citep{Deng:2013aga, Zheng:2014rpa, Macquart:2015uea, Akahori:2016ami, Fujita:2016yve,Shull:2017eow,Ravi:2018ose}, dark matter distribution \citep{Gao:2014iva, Munoz:2016tmg,Wang:2018ydd}, photon mass \citep{Wu:2016brq, Bonetti:2016cpo, Bonetti:2017pym, Wei:2016jgc,Shao:2017tuu,Wei:2018pyh}, the Equivalence Principle \citep{Wei:2015hwd,Tingay:2016tgf,Zhang:2016obv,Yu:2017xbb,Bertolami:2017opd,Yu:2018slt}, the cosmic web \citep{Ravi:2016kfj}, \acrfull{cmb} optical length \citep{Fialkov:2016fjb}, and superconducting cosmic strings \citep{Yu:2014gea, Cao:2018mih}. The efficacy of some of these approaches will, however, rely on the ability to model and remove the \acrshort{frb} host galaxy contribution to the observed \acrshort{dm}. 

The sparsity of observational data has led to the publication of a plethora of progenitor theories in recent years. These can be broadly broken into two groups: those which appeal to astrophysical mechanisms for which there is some empirical evidence (such as plasma physics); and those which appeal to aspects of (mathematical) physics. The latter are more speculative and/or lack any empirical evidence, but nonetheless may be well-motivated from a theoretical standpoint. While the former class may be more appealing to some, here we take an agnostic stance and summarize both sets of possibilities. 

We are aware of recent \acrshort{frb} review articles  \citep{Katz:2016dti,Petroff:2017pny,Romero:2015nec,Popov:2018wei, Katz:2018xiu, Popov:2018hkz}, which have some overlap with this work. However, here we compile a complete catalogue of the \acrshort{frb} progenitor theories published in literature to date. We split the theories into different physical classes, and focus on possible observable signatures and counterparts, with the aim of helping to guide future observing strategies and expedite the process of ruling out theories.

Since the theoretical ideas regarding \acrshortpl{frb} are rapidly evolving, we have launched the contents of this review as an evolving online catalogue at \url{http://frbtheorycat.org}. This online wiki is intended to be a starting point for further discussions amongst theorists, observers and instrument builders regarding \acrshort{frb} theories, observational constraints and future tests. Readers are encouraged to participate in the wiki by signing up and contributing to the discussion. Over the next few years, this wiki will record the zeitgeist of the \acrshort{frb} community. As the wiki stores all changes made to the website, it will also serve as a record of how this scientific community converged on the correct answers.

The organization of this paper is as follows: in Section~\ref{observations}, we summarize the current state of \acrshort{frb} observations and in Section~\ref{modelingredients}, we discuss the basic physical processes that are invoked in \acrshort{frb} theories. These sections are intended as an introduction to the \acrshort{frb} field for new graduate-level researchers. In Section~\ref{theories}, we summarize the proposed \acrshort{frb} theories and their predictions. For the reader's convenience, we offer a tabulated summary as well as a glossary of the acronyms in Section \ref{TableSummary}.

\section{Basic Observational Constraints}
\label{observations}

\subsection{Dispersion Measures}
\label{dm}
One of the primary observables of an \acrshort{frb} is the delay in arrival time between different frequency components of the burst. This delay is proportional to the dispersion measure $\mathrm{DM} = \int n_e \mathrm{d}l$, i.e. the column density of free electrons along the line of sight from the source to the observer. For an extragalactic \acrshort{frb}, the \acrshort{dm} is expected to be the sum of contributions from the Milky Way's disc \citep{Yao:2017} and halo \citep{Dolag:2014bca}, the \acrshort{igm} \citep{McQuinn:2013tmc,Akahori:2016ami}, the disc and halo of the host galaxy \citep{Xu:2015wza,2017ApJ...839L..25Y,Tendulkar:2017vuq}, and the local environment of the \acrshort{frb} progenitor \citep{Lyutikov:2016ueh,Piro:2016aac,Yang:2017vtd}, while the \acrshort{dm} of a galactic \acrshort{frb} must be entirely accounted for by the Milky Way and/or the local environment of the progenitor \citep{Connor:2015era}. Of the \acrshortpl{frb} so far observed at high galactic latitude, most have large \acrshort{dm} that is difficult to account for with gas in the Milky Way, and so point toward and extragalactic origin  \citep{Xu:2015wza}. The presence of excess \acrshort{dm} contributed by H$\alpha$ filaments or HII regions has also been ruled out for most \acrshortpl{frb}.

Since all but one \acrshort{frb} (see Section \ref{curiouscase}) consist of a single pulse with no observable counterparts, they have not been sufficiently localized on sky to be associated with any astronomical object, and thus their distance is unknown. However, if one assumes that the dominant contribution the \acrshort{dm} is due to diffuse gas in the \acrshort{igm}, one can estimate their distance/redshift. These estimates suggest a cosmological origin \citep{Dolag:2014bca,Kulkarni:2014vea,Katz:2015mpa,Caleb:2015uuk,Vedantham:2016tjy,Cao:2017ucx,Niino:2018xze}. As other contributions to the \acrshort{dm} are additive, however, these estimates are always upper limits \citep{Cordes:2016rpr}. If the source resides in a galaxy and/or is surrounded by ejecta/nebula, the \acrshort{dm} will increase from this local contribution \citep{Keane:2012yh,2015RAA....15.1629X,Xu:2015wza, Cordes:2015fua, Connor:2015era}. 

\subsection{Polarization and Rotation Measures}
\label{rm}
Linearly polarized light passing through a plasma with a magnetic field component along the direction of travel undergoes Faraday rotation---where the plane of polarization rotates around the direction of travel by an angle proportional to the rotation measure and wavelength squared. The \acrshort{rm} is defined as $\int B_{||} n_e \mathrm{d}l$, where $B_{||}$ is the component of magnetic field parallel to the direction of travel.

Contributions to the \acrshort{rm} can be from the Milky Way's disc \citep{Oppermann:2011td} and/or halo, the \acrfull{igm} \citep{Akahori:2016ami}, the disc and/or halo of the host galaxy, or the local environment \citep{Connor:2015era,Piro:2016aac,Lyutikov:2016ueh,Michilli:2018zec}. For the handful of \acrshortpl{frb} that have been successfully measured, polarizations are observed to be circular \citep{Petroff:2014taa,Caleb:2018oyn} and/or linear \citep{Masui:2015kmb,Ravi:2016kfj,Petroff:2017pff,Michilli:2018zec,Gajjar:2018bth}. The \acrshortpl{rm} have also been varied---some are very high \citep{Masui:2015kmb, Caleb:2018oyn,Michilli:2018zec} and others are very low \citep{Keane:2016yyk,Petroff:2017pff,Caleb:2018oyn}. The polarizations and \acrshortpl{rm} therefore serve as a useful probe of the magneto-ionic environment associated with the \acrshort{frb}. Together with the \acrshort{dm}, one can also use the \acrshort{rm} to approximate the mean magnetic field strength between a source and an observer.

\subsection{Observed Counterparts / Possible Counterparts}
\label{counterparts}
The presence of a signal unequivocally associated with an \acrshort{frb} will break the viability of certain theories. Here we present the \acrshortpl{frb} for which potential counterparts have been suggested or detected. 

\subsubsection{FRB 150418}
In 2016, the first possible \acrshort{frb} counterpart was claimed \citep{Keane:2016yyk}: a $\sim 6$ day radio transient following FRB 150418 with emission consistent with a \acrfull{sgrb} afterglow \citep{Keane:2016yyk,Zhang:2016mje}. The signal lacks significant high-energy $\gamma$-ray afterglow \citep{Abdalla:2016uep}, and has shown no apparent optical variation \citep{Niino:2018gho}. Though the source was initially thought to be a fading transient, follow up observations \citep{Williams:2016zys} show the emission is consistent with an \acrfull{agn} \citep{Akiyama:2016yho, Williams:2016zys, Vedantham:2016ufj, 2016MNRAS.463L..36B,Giroletti:2016cns, Niino:2018gho}, however the \acrshort{agn} may not be associated with FRB 150418 \citep{Williams:2016zys, Giroletti:2016cns}. The chance of coincidence is arguably low \citep{Johnston:2016mvx, Li:2016vzg}, however the general consensus is that the events are unrelated. 

\subsubsection{FRB 131104}
A relatively long duration ($\sim 100$s) $\gamma$-ray transient, Swift J0644.5--5111 \citep{DeLaunay:2016xpf}, was observed shortly after FRB 131104 \citep{Ravi:2014mma}. The emission is consistent with a \acrfull{grb}. The presence an expected multi-wavelength afterglow was not observed \citep{Shannon:2016rfk}, but arguably may not be observable \citep{Gao:2016lrw}. A high-energy (GeV) $\gamma$-ray afterglow is not affected by environment density, but searches for such signals have yielded null results \citep{Xi:2017vcb}. The lack of counterparts suggests that FRB 131104 is not associated with the \acrshort{grb}-like emission. Further, the signal to noise ratio of the detection is very low, and as such it is generally agreed that the emission is not an \acrshort{frb} counterpart. 

\subsubsection{FRB 121102}
To date, only one reported \acrshort{frb} has unequivocally been associated with counterparts. The repeating FRB 121102 (see Section \ref{curiouscase}) was monitored by the Very Large Array in a fast-dump recording mode \citep{Law:2014ioa}, facilitating the discovery of a persistent radio and optical counterpart, and allowing for its localization on a sub-arcsecond scale \citep{Chatterjee:2017dqg, Tendulkar:2017vuq}. Balmer and [O\small{III}] emission lines appear in the opical spectrum that are similar to extreme emission line galaxies \citep{Tendulkar:2017vuq}. Further observations by Arecibo and the \acrfull{evn} found the signals lie within $40$ pc of each other \citep{Marcote:2017wan}, providing strong evidence that the \acrshort{frb} and radio counterpart are indeed associated. In light of this, FRB 120112's physical origin was swiftly identified: a low-metallicity star-forming dwarf galaxy at redshift $0.19273(8)$ \citep{Tendulkar:2017vuq, Bassa:2017tke}, with the persistent radio counterpart offset from the center of the galaxy \citep{Tendulkar:2017vuq}. 

New evidence suggests that this dwarf galaxy host may not be generic---the recent localizations of non-repeating \acrshortpl{frb} to the outskirts of elliptical galaxies (Bannister et al \emph{in preparation}; private communication) suggests that \acrshortpl{frb} come from heterogeneous environments, possibly suggesting multiple astrophysical formation channels.

\subsection{The Curious Case of FRB 121102}
\label{curiouscase}
As seen in the previous section, FRB 121102 is unique, being the first \acrshort{frb} known to repeat and the only one to have had its location accurately determined. As such, one can study its local environment and narrow down theories describing its source. Though FRB 121102 may be representative of an \acrshort{frb} type distinct from the non-repeaters, it provides the first solid insights into the nature of these strange sources.

The \acrshort{frb} emission has a relatively small \acrshort{dm} and the bursts are consistently $100\%$ linearly polarized with very high and variable Faraday \acrshortpl{rm}:  RM$_{src} = +1.46 \times 10^{5}$ rad/m$^{2}$ and $+1.33 \times 10^{5}$ rad/m$^{2}$ at epochs separated by seven months with narrow temporal structure  ($< 30\mu s$) \citep{Michilli:2018zec}. These measurements have constrained the size of the emitting region to be $\lesssim 10$ km in diameter \citep{Michilli:2018zec}. Making FRB 121102 even more unique is that the \acrshortpl{rm} are 500 times larger than those reported for any other \acrshort{frb} \citep{Michilli:2018zec,Gajjar:2018bth}. Other signatures are that the polarization position angle for FRB 121102 appears to stay constant through each burst and does not display the usual S-shaped curve that is seen in pulsar pulses \citep{Michilli:2018zec}. The intrinsic polarization position angle doesn't change significantly from burst to burst either, with a timescale of months.

\section{Model Ingredients}
\label{modelingredients}
Different models employ different radiation mechanisms to produce the observed radio characteristics of \acrshortpl{frb}, and have a range of corresponding signatures. The mechanisms themselves are based on radiation production, and fall broadly into one of several classes. For the reader's edification and comfort, below we will present each of the overarching radiation mechanisms that astrophysical sources can produce. Additionally, mechanisms to generate the required coherence are given. See \citep{1979rpa..book.....R} for a comprehensive study.

\subsection{Emission Mechanisms}
\label{radiation}
\subsubsection{Bremsstrahlung Radiation}
Translated as ``braking radiation'' or ``deceleration radiation'', Bremsstrahlung occurs when the path of a free electron is bent as it passes a free ion. This deflection causes the electron to decelerate and emit free-free radiation.

\subsubsection{Atomic Electron Transition}
Electrons bound to an atom have discrete energy levels. An electron can either absorb \acrfull{em} radiation and jump to a higher energy state, or emit radiation when dropping to a lower energy state.

\subsubsection{Synchrotron Radiation}
\label{synch}
A charged particle moving in a magnetic field with some component of velocity perpendicular to the field will be forced on a helical trajectory, and emit radiation. The energy and intensity of the emission is determined by the curvature of the charged particle's trajectory. For charged particles that spiral tightly around magnetic field lines, the main contribution to the radiation is from the helix curvature. Such radiation is called: synchrotron (for ultra-relativistic particles), cyclotron (for non-relativistic particles), or gyro-synchrotron (for moderately relativistic particles), and covers a broad range of the electromagnetic spectrum.

\subsubsection{Curvature Radiation}
When the helical path of a charged particle in a magnetic field is stretched out, the curvature of oscillations around the field lines becomes negligible compared to the curvature of the magnetic field line itself, which becomes the dominant contributor to emission. Ultra-relativistic particles that move in a strong magnetic field, for instance, can quickly lose their rotational energy through synchrotron emission, and begin to emit curvature radiation as they stream along the field lines.

\subsubsection{Undulator Radiation}
When electrons pass through an alternating electric field, they are made to oscillate. The radiation is synchrotron for relativistic electrons and cylotron for non-relativistic electrons.

\subsubsection{Inverse Compton Scattering}
When photons are scattered by ultra-energetic electrons, the resultant energy transfer can create photons at significantly higher energies---X-rays or $\gamma$-rays. The spectral energy distribution of the photons is dependent on the original spectrum of the photons, the momentum distribution of the electrons, and the optical depth of interaction. In the case of curvature radiation, the photon momenta are closely aligned with the electron momenta and experience little inverse Compton scattering. However, in the case of synchrotron radiation, the misalignment of photons and electron velocities causes high levels of inverse Compton scattering, and thus X- and $\gamma$-rays are expected counterparts of the radio emission.
	
\subsection{Generating Coherence}
The high energetics and short timescales of \acrshortpl{frb} have led many authors to presume coherent radiation in their progenitor theories. Coherent radiation is emitted by: bunched particles accelerating along \acrshort{em} field lines (also known as an ``antenna'' mechanism \citep{Buschauer:1977}); by masers, in which the emission from simultaneous electron phase transitions is amplified; or by a \acrfull{dsr}, in which particles collectively undergo an atomic transition.

\subsubsection{Bunched Particles}
When particles accelerating along magnetic field lines are bunched into groups such that they have the same oscillatory phase, the radiation emitted is coherent. There are various astrophysical bunching mechanisms possible, described below.

 \paragraph{\textit{Two-Stream Plasma Instability}}
 Consider a plasma in a steady-state force-free magnetosphere with two or more positron and electron streams moving on curved field lines. The velocities of these streams will differ if the net charge density of the plasma varies. The streaming between oppositely charged regions of plasma excites unstable electrostatic plasma waves. The waves trap the particles, forming bunches of electrons and positrons that are accelerated along the magnetic field lines. Streaming can be induced when particles are injected into a plasma.

\paragraph{\textit{Magnetic Reconnection}} A violent disturbance can cause magnetic field lines from different domains to snap and then splice together. Such an event causes particles to bunch in different regions of the magnetic field lines. The relaxation of the magnetic field to a lower energy state accelerates these bunches to relativistic speeds, creating coherent radiation.

\paragraph{\textit{Magnetic Braking}} When ionized material from a differentially rotating body is captured by the body's magnetic field and is propelled into space, the body spins down; an effect known as magnetic braking. The ejected bunches of particles are accelerated radially along the magnetic field lines and emit high-energy photons.

\subsubsection{Masers}
Here we limit our discussion to synchrotron masers \footnote{A synchrotron maser can operate at radio frequencies consistent with \acrshortpl{frb} of both galactic and cosmological origins \citep{Ghisellini:2016jmg}, and can possibly produce the circularly and linearly polarized emission observed in some \acrshortpl{frb} \citep{Long:2018kly}. Curvature masers have been considered as a source of curvature radiation for \acrshortpl{frb}, but are found to be non-viable \citep{Locatelli:2017kfk}.}. Consider the following scenario: if an incoming photon is absorbed by an excited electron, the electron will emit two photons (of the same wavelength) when it drops to its lowest energy state. These two photons can excite two electrons (one each), and four photons will be emitted, and so on. This ``negative absorption'' causes an increase of photons in the radiation field. If the number of excited electrons exceeds the number of electrons in the ground state, population inversion is reached, where the rate of stimulated emission exceeds absorption, and the radiation can be amplified. Let us now consider the interaction between an \acrshort{em} wave traveling parallel to a magnetic field, and an incoming beam of electrons spiraling around the magnetic field lines. Initially the phase of electrons in their orbits around the field lines is random, however if the frequency of the \acrshort{em} wave is close to the electron synchrotron frequency, phase bunching can occur. The bunches of electrons will simultaneously jump to the ground state, and emit photons of the same wavelength to start the chain of stimulated emission described above, amplifying the radiation and thereby forming a maser.

\subsubsection{Dicke's Superradiance}
\label{DSR}
In quantum optics, \acrfull{dsr} is emission from a quantum entanglement of atoms or molecules. When an atom/molecule emits a photon through an transition, it will do so over some defined timescale. When a large collection of atoms/molecules are entangled, a transition in one induces transitions in the others. The atoms will then simultaneously emit photons coherently, at a higher energy over a shorter timescale than if they were not entangled. There is also a characteristic delay of \acrshort{dsr} associated with bulk emission.

\section{Progenitor Theories}
\label{theories}

We now dive into our main agenda for this article: the cataloguing of various progenitor theories and their counterparts for experimental verification or exclusion. 

\subsection{Compact Object Mergers / Interactions}
\label{merger}

\subsubsection{Neutron Star--Neutron Star Mergers / Interactions}
\label{nsmerger}
In this section we consider three theories of a similar kin and then give the counterparts expected in such scenarios.

\paragraph{\textit{Magnetic Braking}} Let us consider the merging of two differentially rotating \acrfullpl{ns} into a uniformly rotating hypermassive \acrshort{ns}. Upon coalescence, the merger spins down and magnetic braking generates coherent radiation. Since the merger rate of \acrshortpl{ns} is consistent with the expected \acrshort{frb} rate \citep{Totani:2013lia}, this model implies that a large fraction of \acrshort{ns} mergers must produce \acrshortpl{frb}. Significant mass ejections are likely to occur during the merger process, which render \acrshortpl{frb} unable to penetrate the ejecta \citep{Hotokezaka:2012ze}. There is a $\sim1$ ms time frame after the maximum rotation speed of the merger is reached and before the release of dynamical ejecta, during which the transmission of a single \acrshort{frb} is possible. Once the ejecta has sufficiently dissipated $\sim 1-10$ years after the merger, pulsar wind nebula energized by the \acrshort{ns} could be a source of repeating \acrshortpl{frb} \citep{Yamasaki:2017hdr} (see Section \ref{SNRep} for further detail.)

\paragraph{\textit{Magnetic Reconnection}} A unipolar induction model has been used to study \acrshortpl{ns} during the final stages of their inspiral \citep{Wang:2016dgs}. As the stars approach each other, a toroidal magnetic field is induced around the star's individual magnetic field lines. The toroidal field strength eventually builds up to rival the poloidal field of the magnetospheres, resulting in magnetic reconnection. After reconnection, the toroidal magnetic field becomes weak again and the process can repeat as the \acrshortpl{ns} continue to spiral inwards. If emission occurs for two orbital periods, a double peaked \acrshort{frb} could be observed, as in FRB 120012 \citep{Champion:2015pmj}. In this theory, the merger product is presumed to be a rapidly spinning \acrfull{bh}.

\paragraph{\textit{Changing Magnetic Flux}} A close encounter between two \acrshortpl{ns} results in the formation of highly elongated orbits, continuously shrinking due to dissipation through \acrfullpl{gw}. As the orbital separation decreases, the changing magnetic flux bunches particles, which then radiate coherently. The \acrshortpl{ns} will approach each other several times, during which their respective magnetospheres will interact, leading to several \acrshortpl{frb} before the final merger \citep{Dokuchaev:2017pkt}. Optimistic predictions indicate that the expected collision rates in galaxy centres are consistent with \acrshort{frb} observations.

Expected counterparts to \acrshort{ns} mergers are \acrshortpl{gw} \citep{TheLIGOScientific:2017qsa}, X-rays \citep{Rees:1976cj, Brown:1995mv,Hansen:2000am, McWilliams:2011zi, Lai:2012qe}, \acrshortpl{sgrb} and afterglows \citep{Paczynski:1986px, Eichler:1989ve, Narayan:1992iy, Dai:2006hj, Zhang:2012wt}, and kilonovae \citep{Li:1998bw, Metzger:2010sy}. The \acrshortpl{sgrb} themselves are not necessarily observable; the direction of curvature radiation doesn't always align with the orbital angular momentum along which \acrshortpl{sgrb} jets are emitted. The short duration \acrshort{grb} afterglow, with its wider opening angle, would be a better signature to search for. In a kilonovae, as the stars coalesce, radioactive elements formed in a rapid neutron capture process (r-process) decay to power optical or near infra-red (NIR) emission. The latter is likely insignificant \citep{Peng:2018ywf}, but optical emission may be observable. The signal is isotropic and detectable for a period lasting days to weeks, making kilonovae good candidates for observational detection of counterparts. It is difficult to distinguish kilonovae from supernovae, but the rapid decay and/or color evolution of kilonovae could help differentiate them \citep{Niino:2014fja}. Note that GW 170817 has been observed with a kilonova and \acrshort{em} counterparts, however follow up searches for \acrshortpl{frb} have yielded null results \citep{2017PASA...34...69A}.

\subsubsection{Neutron Star--Supernova Interactions}
One of the first theories posited that a millisecond pulse, akin to the Lorimer burst, could be formed when a supernova shock interacts with the magnetosphere of a \acrshort{ns} in a giant binary \citep{Egorov:2008yj}. When the shock encounters the \acrshort{ns} magnetosphere, it sweeps out a magnetospheric tail, which triggers reconnection and hence emission. A \acrshort{grb} is expected in such an event, but with a low flux that may be difficult to detect. 

\subsubsection{Neutron Star--White Dwarf Mergers}
Next we consider the interaction between the bipolar magnetic fields of a \acrshort{ns} and a magnetic \acrfull{wd} as a possible origin of repeating \acrshortpl{frb} \citep{Gu:2016ygt}. As the \acrshort{wd} exceeds its Roche lobe, the \acrshort{ns} accretes the infalling matter. Upon their approach, the magnetized materials may trigger magnetic reconnection and emit curvature radiation. In a rapidly rotating \acrshort{ns}, the angular momentum added by accretion is lost to gravitational radiation, but the mass transfer may be violent enough for the angular momentum of the \acrshort{wd} to dominate over the gravitational radiation. In this case, the \acrshort{wd} is kicked away from the \acrshort{ns}, and the process of accretion, and thus magnetic reconnection, may repeat. The timescale of emission is assumed to be the same as that of magnetic reconnection, and the time interval between adjacent bursts is derived from its relationship to the mass transferred by the first burst. There are multiple parameter sets that can describe a repeating \acrshort{frb}, an example of which produces timescales roughly consistent with FRB 121102. Counterparts to this model are not specified, other than to say that possible $\gamma$-ray emission from synchrotron radiation is unlikely detectable.

The emission of a single \acrshort{frb} upon NS-WD coalescence has also been suggested \citep{Liu:2017ctq}. As they merge, magnetic reconnection injects relativistic electrons from the surface of the \acrshort{wd} into the magnetosphere of the \acrshort{ns} to create a burst of coherent curvature emission. The timescale of the burst is assumed to be from the time of electron injection to the formation of the final merged object. It is predicted that the shorter the intrinsic width of the \acrshort{frb}, the higher the flux density. Of the 28 \acrshortpl{frb} analyzed (those available at the time), the pulse widths were broader than expected in the \acrshort{ns}--\acrshort{wd} scenario, but perhaps pulse widths vary more widely between \acrshortpl{frb} due to multipath scattering through the \acrshort{igm} \citep{Cordes:2002wz,Ravi:2017xph}. As more \acrshortpl{frb} are detected and further constraints are placed on pulse widths, this theory can be tested.

\subsubsection{Binary White Dwarf Merger}

It has been proposed that \acrshortpl{frb} may form when a doubly-degenerate binary \acrshort{wd} merger forms a rapidly rotating, magnetized, massive \acrshort{wd} \citep{Kashiyama:2013gza}. The event rate of such a scenario is consistent with that predicted for \acrshortpl{frb}. The rapid rotation of the \acrshort{wd} merger transports, via convection, inner magnetic fields to the polar regions, which greatly enhances the magnetic field strength. The corresponding energy budget has been shown to be sufficient for \acrshort{frb} production. In the polar regions, where the magnetic fields are twisted by differential rotation or magnetic instabilities, reconnection is triggered, and electron bunches are injected into the polar region in a timescale comparable with \acrshortpl{frb}. The electrons are accelerated to relativistic speeds along magnetic field lines, creating curvature radiation. \acrshortpl{wd} transfer angular momentum into the surrounding debris disk, rapidly reducing their rotation speed and hampering multiple \acrshort{frb} events. The \acrshortpl{frb} are predicted to have one of two possible counter parts: X-ray emission, or a Type Ia \acrfull{sn} formed upon the collapse of the massive \acrshort{wd}. Whether \acrshortpl{frb} can penetrate \acrshort{sn} ejecta has, however, been called into question (see Section \ref{SNRep}).

\subsubsection{White Dwarf--Black Hole Mergers}
During the merger of a \acrshort{bh} and a \acrshort{wd}, a transient accretion disk is expected to form around the \acrshort{bh}, which could power a high speed wind around the entire \acrshort{bh}-accretion disk system, forming a corona \citep{Dong:2017mln}. Closed magnetic field lines, emerging continuously between the accretion disk and the corona, are twisted by the turbulence in the system, leading to the formation of rope-like flux structures in the corona \citep{Yuan:2008zv}. When the threshold for mass equilibrium is exceeded, the rope is thrust outward as an episodic jet of relativistic magnetized plasma; a so-called ``magnetic blob''. Before the accretion disk is exhausted, 2--3 magnetic blobs could be ejected at different speeds and will collide at a time after ejection \citep{Li:2018bjq}. The collision causes catastrophic magnetic reconnection, and the release of magnetic energy is propagated through the magnetized cold plasma of the blob, and converted to particle kinetic energy. The resulting synchrotron maser could power a non-repeating \acrshort{frb} \citep{Li:2018bjq}. The expected duration, frequency and energetics in this scenario are consistent with \acrshortpl{frb}, and the event rate of \acrshort{bh}--\acrshort{wd} mergers is compatible with that expected for non-repeating \acrshortpl{frb} \citep{Li:2018bjq}. Note that the accretion disk is advection-dominated. If the disk has a neutron-dominated accretion flow, only a single blob can be ejected within the lifetime of the accretion disk, and thus no collision will take place. X-ray emission from the accretion disk is expected, which will last only as long as the transient disk itself \citep{Li:2018bjq}.

\subsubsection{Neutron Star--Black Hole Mergers}
\label{nsbhmerger}

During the inspiral of a \acrshort{ns}--\acrshort{bh} merger, the magnetic field lines of the \acrshort{ns} may thread around the \acrshort{bh} event horizon in a way similar to a battery powering a circuit \citep{McWilliams:2011zi}---the \acrshort{ns} acts as an external resistor, the magnetic field lines as wires, the magnetospheric $e^-e^+$ plasma providing current, and the \acrshort{bh} provides the power. The charged particles in the \acrshort{ns} magnetosphere are propelled along magnetic field lines by the \acrshort{bh} \citep{ Ardavan:2009zk} and generate radiation akin to \acrshortpl{frb} \citep{Mingarelli:2015bpo}. Under usual circumstances the coalescence of such a system is expected to remain dark; most non-spinning \acrshortpl{bh} will engulf a companion \acrshort{ns} whole, with the disruption radius within the Schwarzchild radius, preventing \acrshort{em} radiation from escaping. In this scenario, however, the battery phase generates high energy X- and $\gamma$-rays. The \acrshort{frb} emission would have a distinct signature, with a precursor and a double peak. The precursor is a rapid increase in luminosity, 20--80\% as bright as the main burst. The double peak consists of a main burst prior to the merger and a post-merger burst, where the former is the peak luminosity, and the latter is from a shock created when the \acrshort{ns}'s magnetic field migrates to the \acrshort{bh}. While the duration and luminosity of \acrshort{ns}--\acrshort{bh} radio bursts have been shown to be consistent with \acrshort{frb} observations, the estimated \acrshort{frb} event rate used at the time was approximately three orders of magnitude greater than the optimistic estimate of the \acrshort{ns}--\acrshort{bh} merger rate \citep{Mingarelli:2015bpo}. A more recent estimate, however, reduces the \acrshort{frb} event rate by an order of magnitude \citep{Champion:2015pmj, 2017AJ....154..117L}, thereby decreasing the discrepancy.

\subsubsection{Pulsar--Black Hole Interactions}

It has been proposed that an enhanced giant pulse generated by a rapidly spun-up \acrshort{ns} near a spinning \acrshort{bh} of mass $M \sim 12M \astrosun$ could produce a non-repeating \acrshort{frb} \citep{Bhattacharyya:2017orq}. A gyroscope is used to model the pulsar's spin-precession, which has been shown to increase rapidly near the event horizon of a \acrfull{kbh} \citep{Chakraborty:2016mhx}. Eventually the spin precession exceeds the pulsar's spin, and the latter can be neglected. As such, the pulsar magnetosphere is essentially rotating around the spin-precession axis. The rapid spin-up causes a giant pulse (see Section \ref{giantpulse}), whose emission is consistent with an \acrshort{frb}. The event rate of \acrshortpl{frb} is also consistent with predictions of this theory. The pulsars considered in this scenario are fairly old (to account for the time taken for the \acrshort{ns} and \acrshort{bh} to merge) and therefore \acrshort{sn} not are not expected. The presence of \acrshort{em} counterparts are uncertain, but \acrshortpl{gw} may be detectable. If two or more bursts are released during the rapid spin-up, the event duration is expected to be $>1$ms for an unresolved burst and a double peaked profile is expected if the burst is partially resolved. The calculations used invoke a point gyroscope, which could affect the results significantly. Also, the influence of the spin-precession and the tidal force of the \acrshort{bh} on the pulsar and its magnetosphere has not been considered. Further investigation into the theory is thus required.

\subsubsection{Kerr-Newman--Black Hole Interactions}
Next we investigate the mergers of \acrfull{knbh} systems---spinning \acrshortpl{bh} that carry a electric charge.
\label{bhmerger}

\paragraph{\textit{Inspiral}} A binary \acrshort{bh} system in which at least one of the spinning \acrshortpl{bh} carries a charge would induce a global magnetic dipole normal to the orbital plane. During inspiral, as the orbital separation decreases, the magnetic flux of the system changes rapidly, which leads to particle bunching and the emission of coherent curvature radiation. And for some minimal values of the charge of the \acrshort{bh},  this scenario could produce an \acrshort{frb} and a \acrshort{sgrb} \citep{Zhang:2016rli}. The detection of both signals could provide a lower limit on the charge, and the non-detection of a \acrshort{sgrb} could provide an upper limit. Lending support to the model itself, the association of a \acrshort{sgrb} with \acrshort{gw} 150914 \citep{Connaughton:2016umz} can be consistently  explained in this scenario \citep{Zhang:2016rli}. However, the estimated merger rate of binary \acrshort{bh} systems is $\sim20$ times smaller than that estimated for \acrshortpl{frb}, and thus the theory can only account for the full population in the unlikely case that only $\lesssim 5\%$ of \acrshortpl{frb} are of cosmological origins.

\paragraph{\textit{Induced Magnetosphere Collapse}}
Magnetospheric instabilities in \acrshortpl{knbh} are studied in \citep{Liu:2016olx}, in which the authors use a test particle to show that if the magnetospheres of \acrshortpl{knbh} have unstable regions, with a certain charge of the particle, and charge and spin of the \acrshort{bh}, the resulting emission is consistent with \acrshort{frb} observations. For merging \acrshort{bh} binaries in which at least one of the \acrshortpl{bh} is Kerr-Newman, instabilities due to tidal forces induce reconnection in the \acrshort{knbh} prior to coalescence. The magnetic field violently reconnects, triggering strong relativistic shock waves through the surrounding plasma to produce curvature radiation. $\gamma$-rays could occur when relativistic particles associated with the shock are cooled through synchrotron radiation and inverse Compton scattering \citep{Gao:2012sq}. The resulting \acrshort{frb} would be a precursor to a \acrshort{gw} event. To differentiate these events from \acrshort{ns} mergers one must note: \acrshortpl{knbh} have no thermal emission because they lack solid surfaces, and are non-pulsating because their magnetic and rotation axes are aligned. In order to observe the \acrshortpl{frb}, the \acrshort{bh} spin must be pointing towards the observer. As with the previous \acrshort{bh}--\acrshort{bh} merger scenario, this model can only account for a sub-population of \acrshortpl{frb}.

\subsection{Collapse of Compact Objects}

\subsubsection{Supramassive Neutron Star to Kerr-Newman Black Hole}
\label{NSKNBH}
Upon the collapse of a supramassive \acrshort{ns} into a \acrshort{bh}, an event horizon will likely form before most of the mass and radiation can escape. Here we specifically consider an isolated and magnetized supramassive rotating \acrshort{ns} that collapses into a Kerr-Newman \acrshort{bh}. By the no-hair theorem, magnetic fields are forbidden from piercing the event horizon, and so the magnetosphere will be left behind. Although applicability of the no-hair theorem to \acrshort{ns} collapse has been questioned \citep{Lyutikov:2011tk}, numerical simulations suggests it holds \citep{Dionysopoulou:2012zv,2012PhRvD..86j4035L,2017MNRAS.469L..31N}. Further, within the same constructs of this \acrshort{frb} model, the no-hair theorem can be avoided \citep{Punsly:2016abn}: if a \acrshort{ns} collapses into a metastable \acrshort{knbh}, its electric discharge can cause the magnetosphere to be shed. Violent magnetic reconnection outside the horizon would then induce a strong magnetic shock wave that moves through the remaining plasma at the speed of light, resulting in curvature emission could produce a non-repeating \acrshort{frb} \citep{Falcke:2013xpa}. The timescales and energetics are consistent with \acrshortpl{frb} \citep{2018ApJ...864..117M}. The relatively clean environment is unsuitable for generating \acrshortpl{grb}, and thus gravitational waves are the only expected \acrshort{frb} counterparts. If, however, the supramassive \acrshort{ns} collapses into a \acrshort{bh} shortly after its birth, an X-ray afterglow and short/long \acrshort{grb} may be observable prior to the \acrshort{frb} \citep{Zhang:2013lta}. In order to attain the required event rate, $\sim 3\%$ of \acrshortpl{ns} must be supramassive, rapidly rotating, and highly magnetized. Whether enough of these \acrshortpl{ns} can form is uncertain and such formation mechanisms are largely unknown.

\subsubsection{Neutron Star to Quark Star}

As a \acrshort{ns} spins down, the reduction in centrifugal forces cause the density in the core to increase to the point where neutrons may split in to their constituent parts---a process known as quark deconfinement. This phase transition from neutrons to a quark-gluon plasma triggers a massive explosion---a quark nova---in which the parent \acrshort{ns} collapses into a quark star \citep{1970PThPh..44..291I,Ouyed:2001ts,Staff:2006kq,Ouyed:2013sra}. The outer layers of the \acrshort{ns} are ejected at relativistic speeds, generating highly unstable rapid neutron-capture (r-process) elements, which undergo a rapid series of $\beta$ decays \citep{Jaikumar:2006qx,Charignon:2011ey}. The electrons emitted from this decay stream into the magnetosphere to generate synchrotron emission. Although the duration of the burst is predicted to be over several seconds, as opposed to the millisecond timescale observed for \acrshortpl{frb}, it has been argued that the use of a de-dispersion process that stacks frequency channels to a common initial time is incorrect, and that the observed pulse is in fact a series of self-similar signals at different frequencies that takes place on a larger timescale \citep{Shand:2015uda}. If this is the case, the observed \acrshort{dm} would be far lower in actuality. The expected counterparts are \acrshortpl{gw} from the explosion and from the quark star oscillation modes \citep{GondekRosinska:2003iy,Flores:2013yqa}, and X- and $\gamma$-ray emission directly from the quark star \citep{Ouyed:2010mm}.

\subsubsection{Dark Matter Induced Neutron Star Collapse}

It is possible that a \acrshort{ns} could capture ambient dark matter particles as they scatter off the \acrshort{ns} nucleons and become gravitationally bound. Once the dark matter particles thermalize to the \acrshort{ns} temperature, they sink to the center of the \acrshort{ns}. Here they accumulate until they reach a critical mass and collapse into a \acrshort{bh} \citep{Bramante:2014zca}. The \acrshort{bh} will then engulf the \acrshort{ns}, ejecting the \acrshort{ns} magnetosphere, causing violent magnetic reconnection. The resultant coherent curvature radiation may be consistent with a single \acrshort{frb} \citep{Fuller:2014rza}. The lifetime of a \acrshort{ns} undergoing dark matter-induced collapse is proportional to the density of dark matter in its local environment. In regions of low dark matter density, \acrshort{ns} lifetimes are of the Hubble scale, however where the dark matter densities are high, the final \acrshort{ns} collapse may be observable today. Origins are thus expected to be central regions of high density galaxies---i.e. massive spirals, early type galaxies, and central cluster galaxies---and the event rate is consistent with that of \acrshortpl{frb} \citep{Champion:2015pmj, 2017AJ....154..117L}. No transient counterparts are predicted, and (because of dark matter annihilation) no galactic center X-ray or $\gamma$-ray excess \citep{Fuller:2014rza}. Whether or not the properties of dark matter are capable of inducing \acrshort{ns} collapse remains speculative \citep{Bramante:2014zca}. If \acrshortpl{ns} in central galaxy regions are found to be older than the expected dark matter-induced collapse time, the rate of dark matter accretion is likely too low to form \acrshortpl{bh}. Further, if one does not see a large population of \acrshortpl{bh} with masses akin to \acrshortpl{ns}, the theory must be ruled out. 

\subsubsection{Collapse of Strange Star Crust}

Charge separation in \acrfullpl{ss} can induce large electric fields emanating from their core, which, through polarization of nearby surrounding hadrons, can lead to the formation of a hadronic crust around the star \citep{Alcock:1986hz}. Should the \acrshort{ss} accrete a sufficient amount of matter, the hadrons in the crust will tunnel across the Coulomb barrier, to the \acrfull{sqm} core, where they too are converted to \acrshort{sqm} \citep{Zhang:2018zdn}. This accretion heats the core, hastens the tunneling process, and eventually and inevitably leads to the collapse of the hadronic crust \citep{Kettner:1994zs, Huang1997}.  As it collapses, the magnetic field lines associated with the crust are dragged with the matter, causing a disruption in the field lines of the \acrshort{ss} core. Thus, via magnetic reconnection, $e^{-}e^{+}$ pairs are accelerated to ultra-relativistic speeds along the magnetic field lines, generating a thin shell of relativistic particles that accelerate around the bare \acrshort{sqm} core to emit curvature radiation \citep{Zhang:2018zdn}. Even a small portion of the magnetic energy held in the polar cap region of the \acrshort{ss} core would be sufficient to power an \acrshort{frb} and the timescales of collapse are consistent with observations. There is little thermal radiation in this scenario, and thus no counterparts at X- or $\gamma$-ray frequencies are likely to be observed.

\subsection{Supernovae Remnants}
\label{snrem}
In \acrshort{frb} theories, pulsars and magnetars immersed in \acrshortpl{sn} have been a popular line of investigation; the former invokes high spin-down energies and the latter strong magnetic fields. The rich ejecta of young magnetars or pulsars can provide the observed \acrshort{dm} and their expanding ejecta could in principle produce the required high \acrshortpl{rm} \citep{Michilli:2018zec}. Appropriate \acrshort{ns} candidates may occur in neighboring galactic centers \citep{Pen:2015ema} and---as is consistent with the localization of FRB 120112---star-forming regions of the host galaxy \citep{Kulkarni:2015sxv, Xu:2016vjj}. 

\subsubsection{Giant Pulses}
\label{giantpulse}
Let us first consider rotationally-powered \acrshortpl{frb} from pulsars.

\paragraph{\textit{Young Rapidly Rotating Pulsars}}
Analogous to models for the Crab pulsar, \acrshortpl{frb} of extragalactic origins may be giant pulses of a young pulsar \citep{Keane:2012yh, Cordes:2015fua, Connor:2015era}. The mechanism for coherence in pulsar emission is currently an open question, however curvature radiation by particle bunching is a strong candidate---consistent with \acrshortpl{frb}---and has been explored in detail \cite{2017arXiv171202702Y}. A specific giant pulse mechanism has been proposed for \acrshortpl{frb}, in which a nearly charge-neutral clump of particles (produced by a two-streaming instability or a bunching instability) is accelerated through the pulsar magnetosphere by some reconnection event. The resultant coherent curvature radiation will be emitted for the duration that the clump remains intact \citep{Cordes:2015fua}. Note that this scenario represents only one possibility. \acrshortpl{frb} are predicted to be repeating and stochastic \citep{Connor:2015era}. A \acrshort{sn} explosion a few years prior to the \acrshort{frb} may be observable \citep{Lyutikov:2016ueh}, however, in contrast to the giant flare model that follows, giant pulses are not expected to have observable higher energy signals \citep{Bilous:2012zw, Mickaliger:2012me, Aliu:2012di, Lyutikov:2016ueh,Lyutikov:2016qio}. The \acrshort{dm}, \acrshort{rm} and polarization of \acrshortpl{frb} in this scenario are owed to the nebula surrounding the pulsar as opposed to the \acrshort{igm}. This places \acrshortpl{frb} at extragalactic (as opposed to cosmological) distances \citep{Connor:2015era}, and thus relaxes the energy requirements. In support of the theory, it has been shown that the high \acrshortpl{rm} can be achieved with a \acrfull{pwn} \citep{2018ApJ...861..150P}. Repetitions of \acrshortpl{frb} would continue as a pulsar spins down. For Galactic pulsars whose ages are older than the spin-down timescale, the radio luminosity either increases or remains nearly constant as the the spin-down luminosity of the pulsar decreases \citep{Szary:2014dia}. For the young pulsars considered here, on the other hand, the spin-down luminosity decreases within a timescale of a few years \citep{Kisaka:2017tbf,Lyutikov:2016ueh}. The giant pulse model therefore depends on the observation of rapid flux decay in \acrshortpl{frb} within a few years \citep{Kisaka:2017tbf}.

\paragraph{\textit{Schwinger Pairs}}
In a pulsar that is born with an extremely high spin and a magnetic field comparable to that of a magnetar, the induced electric field in the magnetosphere may be capable of drawing $e^-e^+$ pairs from the magnetosphere vacuum \citep{Lieu:2016hfw} via the Schwinger mechanism \citep{Schwinger:1951nm}. These so-called Schwinger pairs are then accelerated in opposite directions along the magnetic field lines to neutralize the electric field, causing oscillations in the field about zero. Schwinger pairs are created in the polar cap region, where the electric field is strongest, and the coherent curvature emission escapes along the the open magnetic field lines. The duration over which the pairs are produced corresponds to that of an \acrshort{frb} \citep{Lieu:2016hfw}. A repeating \acrshort{frb} is not expected, unless some event causes the \acrshort{ns} to spin-up to its initial state. Such events are likely quite rare, and are unable to account for the full population of \acrshortpl{frb}.

\paragraph{\textit{Pulsar Wind Bubble}}
Consider a pulsar (\acrshort{ns} or \acrshort{mwd}) within a nebula. The dissipation of spin energy drives a \acrshort{pwn} observable as a shell around the \acrshort{ns}. The plasma wind ceases at the termination shock, where the plasma decelerates to sub-relativistic speeds and forms a wind bubble around the pulsar \citep{Dai:2003km}. A subsequent outburst, possibly triggered by pulsar spin-down or by magnetic dissipation in the magnetosphere \citep{Thompson:1995}, will rapidly decelerate when it impacts the \acrshort{pwn}, triggering a \acrshort{grb}. Energy that is not radiated away by the explosion itself travels outwards at a relativistic speed, causing a highly relativistic shock wave to propagate forward into space \citep{Dai:2003km}. Synchrotron emission is generated \citep{Murase:2016sqo}, however the coherence mechanism to generate \acrshortpl{frb} at this point is unknown. A synchrotron maser might result from the coherently reflected particles in the shock front \citep{Gallant:1992}. This coherence mechanism has been considered \citep{Murase:2016sqo}---as it was originally for the \acrshort{frb}-magnetar case \citep{Lyubarsky:2014jta} (see next section)---however for \acrshortpl{ns} and  \acrshortpl{mwd}, the frequency is likely too low to be consistent with \acrshortpl{frb}. A reverse shock wave may give rise to afterglow \citep{Yuanpei-Yang:2016fim}, however both this afterglow and the \acrshort{grb} formed at the shock front are not expected to be observable. Emission from the \acrshort{pwn} as it expands outwards might be detected in the \acrshort{ns} scenario, but not in the \acrshort{mwd} scenario \citep{Murase:2016sqo}. An \acrshort{sn} explosion a few years prior to the \acrshort{frb} may be observed for either body \citep{Lyutikov:2016ueh}. X-rays may be emitted in some pulses, however no \acrshortpl{frb} have been detected in regions that the \textit{XMM-Newton} X-ray telescope covers \citep{Popov:2016jcu}.

The low efficiency of rotation-powered \acrshortpl{frb} has been called to question \citep{Lyutikov:2017ubv}. The problem might be overcome for \acrshort{sn} with a low ejecta mass; as is the case if the \acrshort{sn} is ultra-stripped or sufficiently young \citep{Kashiyama:2017ehl}, however the locations of such \acrshort{sn} are inconsistent with the host galaxy of FRB 121102 \citep{Metzger:2017wdz}. The association of a persistent radio signal to FRB 121102 presents a larger obstacle: the highest luminosity, to date, observed for a Galatic \acrshort{pwn} is only $2 \times 10^{-6}$ times that of the \acrshort{frb}'s counterpart \citep{Michilli:2018zec}. Further, the spectra of bursts from FRB 121102 are well-modeled by a Gaussian distribution \citep{Law:2017rsz,Scholz:2016rpt}, whereas radio pulsar bursts follow a power law distribution \citep{Kramer:2003xr, Jankowski:2017yje}.

\subsubsection{Giant Flares in Magnetars}
\label{giantflares}

\paragraph{\textit{Magnetar Wind Bubble (Single Flare)}}
One of the first postulations for the Lorimer burst was a magnetar hyperflare \citep{Popov:2007uv}. Since then the idea has been widely considered and built upon \citep{Lyubarsky:2014jta,Murase:2016sqo,Beloborodov:2017juh}. An \acrshort{frb} model analogous to the pulsar wind bubble model above has been proposed \citep{Lyubarsky:2014jta}, with the power now deriving from the magnetic energy of a magnetar, and the shock resulting from a giant flare\footnote{One proposed flare candidate is that created in the Schwinger pair model presented in the previous section \citep{Lieu:2016hfw}} impacting the \acrfull{mwn}. A powerful synchrotron maser consistent with \acrshortpl{frb} is formed at the termination shock, either by magnetic reconnection or a ring-like distribution of gyrating particles at the shock front \citep{Lyubarsky:2014jta,Beloborodov:2017juh}. A defining feature of such emission is a hump in the nebula spectrum near the nebula's self-absorption frequency \citep{Yuanpei-Yang:2016fim}.

\paragraph{\textit{Magnetar Wind Bubble (Clustered Flares)}} The shock front of the nebula in the previous scenario cannot recover fast enough to account for the repeating bursts of FRB 121102 and the energy requirements are excessive \citep{Beloborodov:2017juh}. As such, the theory has been adapted \citep{Beloborodov:2017juh}. A hyper-active magnetar is proposed to produce multiple millisecond flares at different energies close to the magnetar. Such a magnetar is young with a hyper-energetic \acrshort{sn} shell and an ultra-fast rotation period. The multiple flares interact to form a series of shocks before reaching the \acrshort{mwn}. As in the previous section, the \acrshortpl{frb} arise from a synchrotron maser formed by gyrating particles at the shock front. Flares in this scenario arise from ambipolar diffusion in the magnetar core; a process which is then enhanced by the strong magnetic fields associated with the high magnetar spin. The flares will therefore be significantly more energetic than those of usual magnetars. Less active magnetars can emit \acrshortpl{frb} by the same mechanism, but these will be non-repeating. Since repeating \acrshortpl{frb} call for rarer magnetars, their event rate is expected to be lower.

From these models, high-energy \acrshortpl{grb} \citep{Yu:2014lwa, Murase:2016sqo,Lyutikov:2016qio,Metzger:2017wdz,Beloborodov:2017juh} and possibly a coincident optical flash from the explosion \citep{Lyutikov:2016qio} are predicted. Flaring from the reverse shock could lead to additional (lower-energy) $\gamma$-ray emission, and the interaction between the \acrshort{grb} and the ejecta could lead to broadband afterglow emission lasting days to weeks \citep{Kulkarni:2015sxv, Murase:2016sqo, Lyutikov:2016qio}. The quasi-steady nebular emission of the magnetar wind nebula itself may be difficult to detect \citep{Murase:2016sqo, Reynolds:2017hbs}. X-rays are able to penetrate the ejecta, but only on a $\gtrsim 100$ year timescale, and are therefore unlikely to be detected \citep{Margalit:2018bje}. A testable signature is that the persistent variable radio source associated with FRB 121102 is predicted to decay by $\sim 10 \%$ within the next few years \citep{Metzger:2017wdz}.

A magnetar that emits bursts at irregular intervals is a \acrfull{sgr}. Although \acrshortpl{sgr} and \acrshortpl{frb} share similar properties, such as: characteristic timescales, low duty factors and repetition \citep{Popov:2007uv, Popov:2013bia,Thornton:2013iua,Kulkarni:2014vea, Lyubarsky:2014jta, Yu:2014lwa, Pen:2015ema, Katz:2016dti, Katz:2015ltv, Katz:2015mpa, Murase:2016sqo, Wang:2016lhy, Beloborodov:2017juh, Metzger:2017wdz, Lieu:2016hfw}, there is a crucial difference. \acrshortpl{sgr} are observed to be entirely thermal with frequencies above the X-ray range, whereas \acrshortpl{frb} are observed in radio frequencies. Another possible inconsistency is that the Parkes Telescope failed to detect an \acrshort{frb} counterpart to the giant flare of SGR 1806-20: only one of the fifteen \acrshortpl{frb} analyzed has a $\gamma$-ray fluence ratio consistent with the \acrshort{sgr} \citep{Tendulkar:2016diq}. The bursts of FRB 121102 have varied spectral characteristics, which suggests the observed fluence ratio may vary significantly between different magnetars and between bursts from the same magnetar. \acrshortpl{frb} therefore may not be observable for all \acrshortpl{sgr}, which would explain the lack of a detectable radio counterpart in SGR 1806-20 \citep{Tendulkar:2016diq}. Searches for \acrshortpl{grb} associated with \acrshortpl{frb} (repeating and non-repeating) with the \acrfull{flat} have not revealed any results, nor have placed any stringent constraints on the magnetar model \citep{Xi:2017vcb, Zhang:2017ddx}. For details on the optimal observing windows for follow up observations of \acrshortpl{frb} associated with \acrshortpl{sgr}, see \cite{Ravi:2014gxa}.

The flares from young magnetars are consistent with the properties of the Lorimer burst \citep{Popov:2007uv} and with the \cite{Thornton:2013iua} \acrshortpl{frb} \citep{Popov:2013bia}. FRB 110523 is in-keeping with magnetar flares, too \citep{Keane:2016yyk}. Based on the observations of SGR 1806-20, the energy and number of particles ($N \sim\ 10^{52}$) in FRB 121102 are found to be consistent with magnetar ejecta, and thus it is arguably more likely to be powered by magnetic fields than rotational energy \citep{Beloborodov:2017juh}. The host galaxy of FRB 121102 supports the predicted \acrshort{lgrb} and hydrogen-poor \acrfullpl{slsn} formed in the birth of millisecond magnetars \citep{Metzger:2017wdz, Nicholl:2017slv}. The variable radio source associated with FRB 121102 is consistent with the giant flare theory, too: it may be emission directly from the \acrshort{mwn}, the shock interaction between the flare and the \acrshort{mwn}, or afterglow from an off-axis \acrfull{lgrb} (such that only the afterglow is observed) \citep{Metzger:2013cha,Metzger:2017wdz}. Note that for the flare model to be consistent, this emission is expected to decay by $\sim 10 \%$ within the next few years \citep{Metzger:2017wdz}. Constraints on the large, decreasing \acrshort{rm} and required radio transparency for FRB 121102 is consistent with a young $\gtrsim 30-100$ year old magnetar with an expanding magnetized electron-ion nebula, akin to those associated with \acrshortpl{slsn} \citep{Margalit:2018bje}. Such a nebula can also account for the observed properties of the variable counterpart associated with FRB 121102 \citep{Margalit:2018bje}.

The flare theory has been met with various criticisms. The upper limit on \acrshortpl{rm} of giant flares with relativistic outflows \citep{Gaensler:2005iq} is 4 orders of magnitude lower than that observed for FRB 121102 \citep{Michilli:2018zec}. The magnetar wind therefore may not have a large enough magnetic field to account for the \acrshort{rm} without a massive \acrshort{bh} in its vicinity \citep{Michilli:2018zec, Zhang:2018zdn}. Should this be the case, the chances of having a young magnetar ($\sim 30-100$ years) near a massive \acrshort{bh} may be lower than having a regular magnetar \citep{Zhang:2018zdn}. The polarization percentage of pulsar emission has been observed to generally increase with decreasing observing frequency \citep{Morris:1970, Manchester:1971, Manchester:1973, Xilouris:1996,Wang:2014xja}, with some pulsars also having a constant linear polarization percentage below some critical frequency \citep{Manchester:1973}. The variation of \acrshort{rm} is also expected to increase as the distance to the pulsar increases \citep{Mitra:2002ip}. To be consistent with $\sim 100\%$ linear polarization and varying \acrshort{rm} ($\sim 10\%$ over 7 months) of FRB 121102, \acrshortpl{frb} must originate some distance from the surface of magnetar. This presents a conflict \citep{Zhang:2018ueg}: the enormous brightness temperatures of \acrshort{frb} emission requires strong magnetic fields close to the magnetar surface \citep{Kumar:2017yiq,Beloborodov:2017juh}. The age of the magnetar must also fall within a ``Goldilocks Zone'': for the appropriate energy budget, the magnetar cannot be too old, but to penetrate ejecta and avoid \acrshort{dm} variation it cannot be too young \citep{Piro:2016aac, Cao:2017ucx, Kashiyama:2017ehl, Zhang:2017ddx, Yang:2017vtd}.

\subsubsection{Ejecta Penetration}
\label{SNRep}
Many authors have shown that \acrshort{sn} ejecta would be opaque to \acrshort{frb}-like signals and could take up to $100$s of years dissipate sufficiently for GHz radio waves to penetrate \citep{Piro:2016aac, Murase:2016sqo, Murase:2016ysq, Piro:2017tmm, Lieu:2016hfw}.  SN 1986J has been used as an example to show the ejecta would become optically thin $\sim 60-120$ years after the core collapse of the parent body \citep{Bietenholz:2017ezg}. By this time, \acrshort{sn} ejecta would be unable to produce a sufficiently high \acrshort{dm}. This is consistent with findings that \acrshortpl{frb} must pass through the intergalactic medium as well as the \acrshort{sn} ejecta \citep{Katz:2015ltv}, which strongly implies cosmological origins. It has been postulated that as \acrshort{sn} ejecta expands and becomes more diffuse, the \acrshort{dm} of \acrshort{frb} pulses from the same source will vary \citep{Piro:2017tmm}. More complex calculations in a follow-up paper, however, show that the \acrshort{dm} can be constant or even increasing \citep{Piro:2018jpl}. This is consistent with repeating FRB 121102, whose \acrshort{dm} appears constant \citep{Michilli:2018zec}. Further searches may show a non-constant \acrshort{dm} \citep{Yang:2017vtd}, but the \acrshort{sn}-\acrshort{frb} theory may remain viable either way. To avoid the ejecta caveat altogether, one may consider that a pulsar is kicked away from its birth state and that the pulsar wind is generated by dense \acrshort{ism} \citep{Dai:2017bjk}. 

\subsection{Active Galactic Nuclei}
\label{AGN}
Various \acrshort{frb} origins involving \acrshortpl{agn} have been suggested: the interaction of an \acrshort{agn} jet with cavitons, the interaction between an \acrshort{agn} and either a \acrshort{kbh} or a \acrshort{ss} and a scaled down version of an \acrshort{agn} jet. Here we summarize these models.

\subsubsection{AGN Jet Interacting with Cavitons} 
Let us first discuss the formation of \acrshort{agn} jets. Consider a hot accretion disk formed as matter is captured and spirals into a moderately sized \acrshort{bh}. Some of the in-falling gas and dust is confined to the poles and ejected in two relativistic jets \citep{Blandford:1977ds}. Hot gas clouds of varying densities surround the \acrshort{bh}, forming a toroid that extends a few parsecs from the \acrshort{bh}. As the \acrshort{agn} jet interacts with the clouds, it becomes narrowly collimated. The relativistic $e^-e^+$-beam encounters material at the center of the host galaxy, and strong turbulence is produced by plasma instabilities. The total pressure and the ponderomotive force (experienced by a charged particle in an oscillating electric field) cause electrons and ions to separate. These regions, called cavitons, are filled by a strong electrostatic field. Electrons from the beam that pass through the caviton are coherently scattered and emit strongly beamed Bremsstrahlung radiation in pulses, consistent with \acrshortpl{frb} \citep{Romero:2015nec,Vieyro:2017flr}. \acrshortpl{frb} may be single or repeating, with the latter shown to be consistent with FRB 121102 \citep{Vieyro:2017flr}. Radiation might be linearly polarized (as observed in FRB 121102 \citep{Michilli:2018zec,Gajjar:2018bth}, FRB 110523 \citep{Masui:2015kmb}, and FRB 150807 \citep{Ravi:2016kfj}) if there is a local magnetic field, however the $100\%$ polarization degree of FRB 121102 would be difficult to account for in this scenario \citep{Zhang:2018ueg}. The persistent scintillating radio emission from the \acrshort{agn} is an expected counterpart, which agrees with observations of FRB 121102.

\subsubsection{Kerr Black Hole Interacting with AGN}
A repeating \acrshort{frb} could be produced when a \acrshort{kbh} surrounded by highly magnetized plasma interacts with an \acrshort{agn} \citep{DasGupta:2017uac}. Episodic winds from the \acrshort{agn} may prompt the \acrshort{kbh} to intermittently accrete matter. Through the Blandford-Znajek mechanism, in which a magnetic field extracts spin energy from the \acrshort{kbh}, an unsteady bipolar jet is triggered. As the jet travels through the surrounding plasma, shocked shells could create a synchrotron maser consistent with \acrshort{frb} observations \citep{Waxman:2017zme}. The \acrshort{kbh} under consideration may form via a series of events: the collapse of a Wolf-Rayet star (a relatively small helium
star) into a supramassive magnetar that accretes matter until it collapses into a \acrshort{kbh}. Precursors to the \acrshort{frb} may therefore be the \acrshort{sn} Type Ib/c explosion in which the magnetar was born, and the resultant neutrinos and \acrshortpl{gw}. The release of a prompt \acrshort{grb} may also be observable as the magnetar spins down, but only if the axis of the magnetar is aligned with earth. \acrshortpl{gw} will also occur when the magnetar implodes to a quark star. 

\subsubsection{Strange Star Interacting with AGN}
A \acrshort{ss} is made up of approximately the same number of up, down, and strange quarks \citep{1970PThPh..44..291I,1978PhRvD..17.1109F}, with a small number of electrons distributed across the star's surface. Should an \acrshort{agn} wind interact with a \acrshort{ss}, it can induce torsional oscillation of the electron layer relative to the positively charged \acrshort{ss}, which can emit high luminosity GHz radio waves \citep{Mannarelli:2014jfa}, consistent with \acrshortpl{frb} \citep{DasGupta:2017uac}. The sporadic nature of \acrshort{agn} wind would induce a repeating \acrshort{frb}. Persistent \acrshortpl{gw} are expected from the \acrshort{ss} due to its r-mode instability \citep{Andersson:2001ev}. If the \acrshort{ss} is the result of a spinning down magnetar, neutrinos and a \acrshort{gw} could be released  when the magnetar collapses, however this emission need not be close in time to the interaction of the \acrshort{ss} with the \acrshort{agn}, making it difficult to draw any associations.

\subsubsection{AGN-like Wandering Beams}
\acrshortpl{frb} may be formed by a scaled down version of an \acrshort{agn} \citep{Katz:2017ube}. The jet formation and beaming mechanism is as in the \acrshort{agn} scenario, but the \acrshort{bh} under consideration has a mass lower than the \acrshortpl{smbh} of \acrshortpl{agn}. If the moderately sized \acrshort{bh} is set in a turbulent medium, such as a giant molecular cloud in a starburst galaxy, the angular momentum axis of the \acrshort{bh} may be large, and the narrowly collimated beams will randomly change directions. When a beam sweeps across an observers line of sight, it may be observable as an \acrshort{frb}. There will be a persistent variable radio signal as in an \acrshort{agn}, and very soft X-ray/extreme UV emission from the accretion disk of the \acrshort{bh}. The latter would be strongly absorbed in the Galactic plane, and thus only observable for \acrshortpl{frb} at high Galactic latitudes.

\subsection{Collisions and Close Encounters}

\subsubsection{Neutron Stars and Small Bodies}
\label{CollNSAst}
Next we investigate a class of progenitor theories based on interactions between a \acrshort{ns} and other gravitationally bound objects.

\paragraph{\textit{Comet/Asteroid Captured by a NS}}
A small body, such as a comet or asteroid, captured by the gravitational potential of a \acrshort{ns} will free-fall toward it, becoming radially elongated until it exceeds its Roche limit and breaks apart \citep{Colgate:1981, Katz:1994tt}. The fragments are compressed by the gravitational acceleration and the magnetic field of the \acrshort{ns}, resulting in leading and lagging portions with the same velocity. In order to nullify the effects evaporation and ionization may have, the body must have a sufficiently large mass and shear, and is thus predicted to be of Fe-Ni composition \citep{Cordes:2006ue}. Infalling matter would be confined to the poles of the \acrshort{ns} by strong magnetic stresses, creating an accretion column \citep{Geng:2015vza}. If the accretion column travels through a region where electrostatic equilibrium has been disturbed, particles are accelerated to yield $\gamma$-ray emission. When the matter impacts the \acrshort{ns}, an expanding plasmoid fireball will be launched along the magnetic field lines. Magnetic reconnection at the collision site accelerates $e^-e^+$ pairs within the plasma-fan to ultra-relativistic speeds. The resulting coherent curvature emission is consistent with a non-repeating \acrshort{frb} \citep{Geng:2015vza,Huang:2015peq}. The event rate associated with such a theory has been shown to be consistent with the other predictions \citep{Thornton:2013iua} and, notably, the impact timescale between the leading and tailing fragments is roughly consistent with the brevity of \acrshort{frb} signals. The model predicts X-ray emission and $\gamma$-ray emission from inverse Compton scattering, however these are probably too faint to be observed \citep{Geng:2015vza}.

\paragraph{\textit{Pulsar Travelling Though Asteroid Belt}}
Consider an asteroid belt surrounding a star. If a pulsar passes through this system, it is likely to encounter multiple asteroids. When this happens, charged particles may be stripped from the asteroidal surface into the \acrshort{ns} magnetosphere, where they are accelerated to ultra-relativistic speeds. The resulting coherent curvature radiation is consistent with \acrshort{frb} properties \citep{Dai:2016qem}. Further, the time between edge on collisions within the asteroid belt is consistent with the time between the signals of FRB 121102. If the pulsar is in a binary system with the star that hosts the asteroid belt, it could pass through the belt multiple times \citep{Dai:2016qem, Bagchi:2017tzi}, creating a series of bursts twice during each orbital period \citep{Bagchi:2017tzi}.

\paragraph{\textit{Body Orbiting Pulsar}} If an orbiting body is massive enough to survive a close encounter without evaporation or breaking up (such as a planet or \acrshort{wd}) \citep{Kotera:2016lwj}, the highly magnetized pulsar wind will induce an \acrshort{em} field around the body. In this situation, Alfv\'{e}n wings are created as the pulsar wind combs the field lines from the nearest pole of the orbiting body and into space. The Alfv\'{e}n wings destabilize the plasma near the body's surface to excite coherent emission. Far from the pulsar companion, the emission is convected with the wind traveling relativistically along the Alfv\'{e}n wings to form a synchrotron maser, whose emission is consistent with \acrshortpl{frb} \citep{Mottez:2014awa}. The emission is only observable when the companion is aligned between the pulsar and Earth, and thus should repeat periodically---a feature yet to be observed for \acrshortpl{frb}. The signal would be composed of one to four peaks, a few milliseconds each, with an event duration less than a few seconds. No emission counterparts are expected, as synchrotron emission from a hot plasma component would be incoherent and thus too weak.

\subsubsection{Collisions Between Neutron Stars and Primordial Black Holes}
\label{NSBH}

\acrshortpl{frb} may result from interactions between \acrshortpl{ns} and \acrfullpl{pbh} \citep{Abramowicz:2017zbp}. As a \acrshort{pbh} passes through a \acrshort{ns}, the gravitational drag from the dense \acrshort{ns} matter causes the \acrshort{pbh} to slow down. The \acrshort{pbh} will pass through the middle of the \acrshort{ns} and, after losing sufficient kinetic energy, will be pulled back. The \acrshort{pbh} will oscillate a few times before settling at the center of the \acrshort{ns}. Here, the \acrshort{pbh} will begin to accrete the \acrshort{ns} until it is swallowed, causing the \acrshort{ns} magnetosphere to be shed. The resulting magnetic reconnection releases an \acrshort{frb}. A repeating \acrshort{frb} may also be accounted for in this scenario: a small \acrshort{pbh} will take longer to accrete the \acrshort{ns}; as the \acrshort{ns} is gradually consumed, multiple bundles of magnetic field lines within the \acrshort{ns} may be reconfigured, causing multiple bursts. \acrshortpl{gw} are expected counterparts, but may not be detectable at cosmological distances. The model can account multiple peaks, polarized emission and Faraday rotation.

\subsubsection{Interactions Between Axions and Compact Bodies} 
\label{AxSNS}

Several progenitor theories have been proposed, which involve axion clusters or clouds coming into close contact with a highly magnetized object. Axions are a prominent candidate for dark matter, and it is expected that they would evolve into clumps of stellar mass, known as axion clouds and stars. At this stage it is worth noting that axion star-\acrshort{frb} models rely on the assumption that axions form a large fraction of the total dark matter component \citep{Eby:2014fya}.

\paragraph{\textit{Axion Star and Neutron Star}}
In the presence of a magnetic field, axions have been shown to produce radiation by generating an oscillating electric field, causing nearby electrons to radiate coherently. The radiation produced when an axion star collides with a \acrshort{ns} has been shown to be consistent with non-repeating \acrshortpl{frb} \citep{Iwazaki:2014wta,Iwazaki:2014wka,Iwazaki:2015zpb}. As the axion star moves through the magnetosphere of the \acrshort{ns}, a time-dependent electric dipole moment is induced, forcing free electrons above the surface of the \acrshort{ns} to oscillate harmonically. This generates coherent radiation with a frequency determined by the axion mass \citep{Iwazaki:2014wta}---an effect which could be even larger if one considers the electric dipole moment induced in the neutrons interior to the \acrshort{ns} \citep{Raby:2016deh}. The theory is shown to be robust to the effects of tidal disruption \citep{Iwazaki:2015zpb}, however this has been disputed \citep{Pshirkov:2016bjr}. A defining feature of the model is that the intrinsic \acrshort{frb} emission frequency is finite, and the observed spectral broadening is due to thermal Doppler effects \citep{Iwazaki:2017rtb}. The emission is also expected to be circularly polarized \citep{Iwazaki:2014wta}. A two-component profile may be observed if the axion star collides with a binary \acrshort{ns} system \citep{Iwazaki:2015zpb}. No counterparts are expected. 

\paragraph{\textit{Axion Star and Black Hole}}
If an axion star were captured by a \acrshort{bh} with a strongly magnetized accretion disk, the axion star's orbit will lead it to approach and impact the accretion disk several times at different locations. The electric field induced by the axion star passing through a strong magnetic field will result in the coherent oscillation of surrounding electrons. In this scenario the frequency of the radiation will depend on the velocity of the accretion disk at the point of impact. In this way, the variation in central burst frequencies of FRB 121102 can be explained. The axion star will likely make several impacts before evaporating or eventually being absorbed by the \acrshort{bh} \citep{Iwazaki:2017rtb}. As in the previous scenario, the emission would be be circularly polarized, and no counterparts are expected. 

\paragraph{\textit{Induced Collapse of Axion Clumps by a Highly Magnetic Compact Object}}
Axion clumps with masses below the stellar range, known as Axion Bose Clusters or ``miniclusters'', have been considered as \acrshort{frb} progenitors \citep{Tkachev:2014dpa}. In the strong magnetic field of a compact object, an instability may arise in a minicluster, causing it to explosively decay into photons via a synchrotron maser mechanism \citep{Tkachev:1986tr}. The predicted emission timescale, the energetics, luminosities, and event rate are in-keeping with \acrshort{frb} observations.

\paragraph{\textit{Superradiant Axion Cloud and Spinning Black Hole}}
Spinning \acrshortpl{bh} have superradiant instabilities, and thus may be surrounded by a dense superradiant axion cloud. Similarly to how a laser can be generated by stimulated axion decay in dense axion clusters \citep{Tkachev:1987cd,Kephart:1994uy}, a laser can be triggered in superradiant axion clusters. Such is known as a \acrfull{blast} \citep{Rosa:2017ury}. For a \acrshort{blast}'s emission to be consistent with \acrshort{frb} observations, the required mass dictates that the \acrshortpl{bh} be primordial. However, because \acrshortpl{pbh} form when over densities of gas collapse \citep{GarciaBellido:1996qt}, they do not have spin, and are unlikely to spin up via accretion \citep{Ali-Haimoud:2016mbv}. The merging of two \acrshortpl{pbh} is thus considered for \acrshortpl{frb}, where the required spin and resultant superradiant instabilities can be induced \citep{Rosa:2017ury}. Repeating bursts could occur: the \acrshort{blast} will form a photon plasma that blocks axion decay and thus halts lasing until $e^-e^+$ annihilation reduces the plasma density, and the process can restart \citep{Rosa:2017ury}. Observational counterparts could be associated with $e^-e^+$ annihilation and/or positronium (a bound particle of an $e^-$ and $e^+$), though these are not specified. \acrshortpl{gw} are also expected. 

\paragraph{\textit{Axion Quark Nugget and Neutron Star}}
In close analogy with the \acrfull{aqn} mechanism for generating solar nano flares  \citep{Zhitnitsky:2017rop}, an \acrshort{aqn} falling through an opportunely complicated region in a \acrshort{ns} magnetosphere may be able to produce sufficient magnetic energy to power \acrshortpl{frb}. Shock waves caused by the infalling \acrshort{aqn} would trigger magnetic reconnection, and produce a giant flare \citep{vanWaerbeke:2018nyj}. The event rate is consistent with observations by \citep{Connor:2015era}, but the emission timescale ($\sim 10-100$ ms) is larger than what is observed \citep{vanWaerbeke:2018nyj}. This discrepancy can be accounted for if the beam moves across the sky, allowing us only a glimpse of the emission. A curvature radiation mechanism is invoked, which predicts a maximum cut-off frequency at infra-red wavelengths; observed counterparts with higher frequencies would invalidate the \acrshort{aqn}-\acrshort{frb} model. Given the random nature of these events, repeating \acrshortpl{frb} would be non-periodic. A correlation between the total energy and duration of the flare is predicted, however because only a fraction of the entire beam would be observed, this relationship would be difficult to verify \citep{vanWaerbeke:2018nyj}.

\subsection{Other Models}

\subsubsection{Starquake-Induced Repeaters}
The starquakes of a pulsar \citep{Ruderman:1969, Baym:1969,Anderson:1975zze} have been considered as a source of repeating \acrshortpl{frb} \citep{Wang:2017agh}. The bursts of FRB 121102 are consistent with the aftershock sequence of an earthquake, where the burst's time-decaying rate of seismicity falls within the typical values of earthquakes. They also show that the burst energy distribution of FRB 121102 has a power law form, much like that of the Gutenberg-Richter law of earthquakes. Further, the waiting time of bursts has a Gaussian distribution; another characteristic feature of earthquakes. Starquakes are poorly understood, limiting the testability of this theory. They may be associated with \acrshortpl{sgr} \citep{Cheng:1996,Xu:2006mp}, which offers counterparts for which to search.

\subsubsection{Variable Stars}
Variable stars may be a source of \acrshortpl{frb} \citep{Song:2017yel}, where synchrotron emission is generated by an astrophysical undulator. The model assumes the existence of a weak, axial-symmetric magnetic field some distance from a variable star. The emission frequency will vary relative to the observer, due to the change in opening angles between the observer and direction of emission as the star rotates. This results in a \acrshort{dm} consistent with \acrshortpl{frb}. Multiple peaks are possible in this scenario. For this model to hold, one must observe a positive frequency sweep ahead of a negative frequency sweep.

\subsubsection{Lightning in Pulsars}

Akin to \acrshortpl{sgrb} powered by the release magnetic energy stored in magnetars, \acrshortpl{frb} may be powered by the release of electrostatic energy stored in pulsars \citep{Katz:2017tcd}. Provided there are regions of magnetospheric plasma with distinct energy, separated by a vacuum gap, the discharge of such energy could manifest as ``pulsar lighting'',  analogous to the flow of current in the atmosphere when lightning strikes \citep{Melrose:2016kaf}. This intense, rapidly varying, electric field in the gaps would accelerate electrons and positrons in the magnetosphere, producing coherent curvature radiation observable as an \acrshort{frb}. The large variation of FRB 121102 burst widths, and hence the variation of spectra fluences and frequencies, may be due to scintillation of pulsar lightning.

\subsubsection{Wandering Pulsar Beam}

The next model assumes the presence of a steady beam of pulsar emission whose direction randomly changes. If this beam sweeps across the line of sight of an observer, it may be observable as an \acrshort{frb} \citep{Katz:2016zxi}. The duration of an \acrshort{frb} depends on the speed at which the beam moves across the sky, and hence a wandering beam mitigates the enormous power and high spin-down requirements of giant pulse and flare models (Section \ref{snrem}). This scenario can also consistently explain two pairs of possibly distinct radio bursts detected in FRB 121102 \citep{Scholz:2017kwy, Hardy:2017djf}. Even with the random walk of the beam, with enough observations one might be able to tease out the periodicity of the pulsar. Details about an emission mechanism or possible counterparts are not specified.

\subsubsection{Tiny Electromagnetic Explosions}

The collision of two relativistic macroscopic dipoles that form around the time of cosmic electroweak symmetry-breaking could cause a tiny\footnote{The initial wavelength of the explosion is narrower than the wavelength of the radiation observed so the energy travels outwards in a very thin shell, and the surrounding charged particles are deflected (as opposed to reflected) by the magnetic field embedded in the expanding shell.} explosion \citep{Thompson:2017wbo}. To retain the supporting electric field for cosmological timescales, these field structures must be superconducting, and are thus dubbed \acrfullpl{lsd}\footnote{``Large'' is of course relative, and is used because the dipoles are macroscopic.} \citep{Thompson:2017inw,Thompson:2017wbo}. The expanding relativistic magnetized shell from the explosion couples efficiently to a low-frequency, strong, superluminal \acrshort{em} wave in the surrounding plasma, allowing the emission to escape. Three emission mechanisms are possible: the reflection of an ambient static magnetic field by the conducting surface of the shell; direct linear conversion of the magnetic field in the shell; and the excitation of an \acrshort{em} wave if the surface of the shell becomes corrugated via the reconnection of the ejected magnetic field with the ambient magnetic field. The deceleration of the magnetic shell causes a higher frequency radio pulse and the thermal part of the explosion radiates $\gamma$-rays, however the latter are not expected to be detectable \citep{Thompson:2017wbo}. The model accounts for repeating and non-repeating \acrshortpl{frb} and for their observed differences in linear polarizations and \acrshortpl{rm} \citep{Thompson:2017inw,Thompson:2017wbo}---the hydromagnetic drag on \acrshortpl{lsd} is weak in the \acrshort{ism}, and strong in high-density environments. As such, a slowly accreting \acrshort{smbh} may capture \acrshortpl{lsd} and group them in gravitationally bound cusps, within which the \acrshortpl{lsd} collide and create repeating \acrshortpl{frb}. Here the high density plasma accounts for the high linear polarization and high \acrshort{rm} observed in FRB 121102 \citep{Michilli:2018zec}. The opposite is true for \acrshortpl{lsd} far from the \acrshort{smbh}; where the observed \acrshort{rm} is low and signals are non-repeating, such as in FRB 150215 \citep{Ravi:2016kfj}, collisions are expected to take place in dark matter halos of galaxies.

\subsubsection{White Hole Explosions}

Should a collapsing star reach the Planck density to become a Planck star, it will cease to collapse further \citep{Goswami:2005fu} and will explode outwards (or bounce) to form a \acrfull{wh} \citep{Hawking:1974rv}. Due to their age, \acrshortpl{pbh} or Planck stars are the strongest candidates to form \acrshortpl{wh} which may be observable today \citep{Carr:1975qj, Rovelli:2014cta}, and the energy they release is consistent with \acrshortpl{frb} \citep{Barrau:2018kyv}. A single \acrshort{frb} is expected, accompanied by an infra-red signal with a wave length on the order of the exploding star and $\gamma$-rays characterized by the material expelled in the explosion \citep{Barrau:2014yka}.

\subsubsection{Neutron Star Combing}
\label{comb}
Cosmic combing is the process in which the field lines of a \acrshort{ns}'s magnetosphere are swept out in a stream by a strong plasma. The effect is caused by ram pressure: the bulk resistance of a fluid acting on an object. When this pressure is greater than the magnetic field pressure, the drag will comb the magnetic field in a different direction, causing reconnection with emission consistent with an \acrshort{frb} \citep{2017arXiv171202702Y, Zhang:2017zse, Zhang:2018ueg}. Combing may occur in a variety of situations, such as: a \acrshort{grb}, a \acrshort{sn}, an \acrshort{agn} flare, or a stellar flare. As such, it is a difficult theory to test in general, however a specific scenario has been considered: the combing of a pulsar by an accreting \acrshort{smbh} \citep{Zhang:2018ueg}. An \acrshort{frb} would be observable for half of the pulsar's orbital period around the \acrshort{smbh}, implying the signal is periodic (as is suggested in \citep{Scholz:2016rpt,Price:2018ezk}). This periodicity would not be perfect---the \acrshort{smbh} wind that initiates \acrshortpl{frb} is variable and thus \acrshort{frb} signals are sporadic. The \acrshort{rm} should vary with orbital periodicity, but this would be more difficult to confirm given the sporadic \acrshort{frb} emission. Finally, the polarization angle of each burst within an orbital period would vary depending on the phase of the pulsar's orbit, and should be correlated with the varying \acrshort{rm}.
 
 \subsubsection{Neutral Strings}
 Nambu-Goto (infinitely thin, idealized) strings generically form cusps---portions of the string which fold back onto themselves and move at the speed of light. The cusps decay to form a beam of coherent radiation \citep{Brandenberger:1986vj}, where the emission can ostensibly be of any energy and frequency range. As such, cusp decay has been considered as an \acrshort{frb} origin \citep{Brandenberger:2017uwo}. The event rate, timescale, and flux are shown to be consistent with \acrshort{frb} data, however the relativistic effects on the cusp shape where not considered. It has been argued that by taking this into account, the consistency of the theory may break down \citep{Costa:2018zwb}.
 
\subsubsection{Superconducting Strings}
A cosmic string becomes superconducting when coupled with electromagnetism; achievable through the unbroken symmetry of an extra Higgs field in the formation of the string \citep{Witten:1984eb}. Various mechanisms have been considered in which superconducting cosmic strings may produce an \acrshort{frb}, such as: string oscillations \citep{Vachaspati:2008su}, the collisions of string structures (cusps and kinks) \citep{Cai:2012zd, Ye:2017lqn}, and the interaction of a current-carrying loop in the magnetic field of a galaxy \citep{Yu:2014gea}. In the last scenario listed, the event rate of \acrshortpl{frb} indicates a loop size consistent with strings formed during the radiation era \citep{Yu:2014gea}. The emission from superconducting cosmic strings is linearly polarized---an intrinsic signature that is independent of frequency and is not affected by polarization via the \acrshort{ism} \citep{Cai:2011bi}.  Expected counterparts are other \acrshort{em} counterparts---specifically \acrshortpl{grb} \citep{Brandenberger:1993hw}, cosmic rays \citep{MacGibbon:1992ug}, and neutrinos \citep{MacGibbon:1989kk}---and \acrshortpl{gw}.

\subsubsection{Dicke's Superradiance in Galaxies}
Postulated in 1953 \citep{Dicke:1954zz} and first detected in 1973 \citep{DSR:1973}, \acrshort{dsr} has been invoked as one of the few ``microphysical'' \acrshort{frb} models \citep{Houde:2017pni}. The aim is to explain \acrshortpl{frb} using the atomic interactions in galaxies.
\acrshort{dsr} can occur in astrophysical settings \citep{DSR:2016}, provided: the collection of atoms is inverted (to wit, a majority of atoms exist in higher excited states than the minority); the velocity coherence is high; and the non-coherent relaxation mechanisms occur on a timescale larger than the delay time. If one models the \acrshort{ism} as a cylinder of atoms, the predicted \acrshort{dsr} emission power and timescale can fit \acrshort{frb} data \citep{Houde:2017pni}. This is because the coherent behavior of the \acrshort{dsr} atoms has a timescale which scales as $\tau \propto N^{-1}$ and an intensity which scales as $I \propto N^2$, where $N$ is the number of entangled molecules. The \acrshortpl{dm} associated with \acrshortpl{frb} fits well with the \acrshort{ism} required for \acrshort{dsr} to occur. The \acrshortpl{frb} arising from \acrshort{dsr} may be repeating; if a collection of molecules has \acrshort{dsr} triggered at the  same time, the intrinsic variation in the \acrshort{dsr} timescale and time delay would give the observation of bursts at different times \citep{Houde:2017pni}. The variation is because the time delay is an expectation value, and the collection of molecules being ionized at the same time is due to the entanglement, which also causes a differential in emission time. This process can happen repeatedly as population inversion will be non-inverted but swiftly restored via the \acrshort{ism}, which will drive more \acrshort{frb} pulses, and so on. The flux distribution of such a setup can be matched to FRB 121102. 

\subsubsection{Alien Light Sails}
Extragalactic, artificial beam-powered light sails have been proposed as an \acrshort{frb} origin \citep{Lingam:2017fwa}, however such a concept is highly speculative, and cannot be tested nor can it provide predictions.

\subsection{Theories That Have Been Ruled Out} 
After all the speculation put forth by the community, some models have already been ruled out. Below, for the sake of completeness, we present a brief outline of the relevant conversations regarding these models. 

\subsubsection{Stellar Coronae}
One potentially promising theory for \acrshortpl{frb} of Galactic origins was flare stars. The theory seemed fitting because dwarf stars have been observed to produce bursts of coherent radiation on short timescales ($< 5$ ms) \citep{Lang:1983, Lang:1986,Bastian:1990} and have been observed within \acrshort{frb} fields \citep{Loeb:2013wna, Maoz:2015xpa}. A cyclotron maser in the lower part of the stellar corona could produce a flare consistent with \acrshort{frb} observations \citep{Loeb:2013wna}, where the large \acrshort{dm} and pulse smearing could be attributed to the corona plasma. Free-free absorption that occurs in the corona, however, presents a problem: a radio signal from the lower corona with the required \acrshort{dm} may be unobservable unless the corona is infeasibly extended or hot \citep{Luan:2014iea}. Further, the plasma density required for the \acrshort{dm} is arguably too high to produce the frequency dependence on the pulse arrival times observed for \acrshortpl{frb}. In defense of the theory, observations by \cite{Dennison:2014vea, Tuntsov:2014uta} show high flare temperatures capable of mitigating significant free-free absorption \citep{Maoz:2015xpa}. Further, if frequency drifts occur in flares (as observed in \citep{Osten:2007fp}), the measured dispersion relationship for \acrshortpl{frb} may be possible. Doubt was then cast on the theory again when it was shown that the brightness temperature of \acrshortpl{frb} could not escape plasma as dense as the \acrshort{dm} demands \citep{Lyubarsky:2015dcj}. The stellar flare theory for \acrshort{frb} origins was likely put to rest when constraints by \cite{Dennison:2014vea, Tuntsov:2014uta, Katz:2015mpa} were tightened to show the density and expanse of the dispersing plasma required to produce \acrshortpl{frb} differs from those of stellar coronae by at least an order of magnitude \citep{Masui:2015kmb}.

\subsubsection{Annihilating Mini Black Holes}
When a \acrshort{bh} evaporates to some critical mass, a fireball of $e^-e^+$ pairs can be created. The relativistic pairs expand into the magnetic field of the surrounding \acrshort{ism}, which, for a \acrshort{bh} of mass $M_{BH} < 10^{13}$ kg, could produce emission consistent with an \acrshort{frb} \citep{Keane:2012yh}. The inferred distance for the Lorimer burst in this scenario, however, is calculated to be $\lesssim 20$ kpc. This would place the source within our galaxy, and thus the theory is rendered void.

\section{Conclusion}
In this review article, we have catalogued a collection of postulated \acrshort{frb} models and have attempted to provide the reader with a general overview of the ongoing research in \acrshort{frb} model-building. The theories vary in their explanatory power, testability, and sometimes in their ``exoticness''. Given both the number of theories and the range of physics used as a foundational framework, it seems like an ideal time to take stock of the theoretical work produced. The small data set and lack of observational counterparts means most of the theoretical work is phenomenological. 

\subsection{Future Observational Constraints}
With a considerable increase in radio data predicted from telescopes such as \acrshort{chime}, \acrshort{lofar}, \acrshort{hirax}, \acrshort{ska}, \acrshort{askap}, \acrshort{utmost}, \acrshort{fast} and \acrshort{apertif} installed in the \acrshort{wsrt}, as well as in other frequency bands by telescopes such as the \acrshort{flat}, \acrshort{gbm} and \acrshort{bat}, it is hoped that constraints on many of the models discussed here will be able to be rule out---and even favor---certain approaches. Here we discuss some of the major ways observations can constrain theory. The constraints are divided into two broad sections and we discuss each in succession.

\subsubsection{Astrophysical Formation Channels}
From an observational perspective, the strongest constraints on the astrophysical channel of \acrshort{frb} formation would likely come from arcsecond-precision localization of \acrshortpl{frb}, leading to an understanding of their environments and their progenitor populations. Historically, this mirrors the expansion of our understanding of long and short \acrshortpl{grb} after the first afterglow localizations with Beppo-SAX. This would provide the least controvertible evidence of any possible distinction between repeating and non-repeating \acrshortpl{frb}---different burst morphologies or statistical rate calculations can be fraught with biases and complications. Current and upcoming \acrshort{frb} search projects with the \acrshort{askap}, \acrshort{hirax}, \acrshort{utmost}-2D and \acrshort{chime}-outriggers promise to provide a large sample of well-localized \acrshortpl{frb} with identifications of their host galaxy environments. The environments of \acrshortpl{frb} will also be constrained by observations of rotation measures, temporal variations in rotation and local dispersion measures. Apart from the environments, the host identification will provide a robust distance and energy scaling for the emission mechanism.  

\subsubsection{Emission Mechanisms}
Most theories expect that the radio emission is a small fraction of energy budget. Detecting or constraining multi-wavelength or multi-messenger prompt counterparts would place constraints on the emission mechanism. The challenge is that the sensitivity of wide-field of view $\gamma$-ray, X-ray, and optical/infrared telescopes is much lower (in $ \nu L_\nu$) compared to that of radio telescopes. The nearest and brightest \acrshortpl{frb} will likely lead to the best constraints and are most likely to have detectable prompt counterparts at other wavelengths. The combination of very wide-field radio telescopes at the GHz frequency band and the all-sky X-ray and $\gamma$-ray burst monitors may lead to strong constraints, albeit on a small number of bright \acrshortpl{frb}. The polarization and temporal characteristics of \acrshortpl{frb}---fraction of linear/circular polarization, change in polarization angle would also constrain the emission mechanisms of \acrshortpl{frb}. Repetition statistics could give phenomenological clues about the energy source and trigger of the emission mechanisms.

Unsurprisingly, we cannot predict the future of \acrshortpl{frb}, how the mystery will be solved and how astronomers will converge on to the answers, but the field certainly will be exciting for many years ahead.

\section{Acknowledgements}

 We would like to thank Renato Costa for many useful discussions during the early part of this work. We would like to thank Bryan Gaensler for many useful discussions about radio astronomy, and for introducing the UCT group to Fast Radio Bursts in the first place. We would like to thank the anonymous referee for their valuable feedback. E. Platts is supported by a PhD fellowship from the South African National Institute for Theoretical Physics (NITheP). A. Walters acknowledges support from the National Research Foundation of South Africa (NRF) [grant numbers 105925, 110984, 109577], and thanks A. Weltman for hosting visits to UCT. S. Tendulkar acknowledges support from a McGill Astrophysics postdoctoral fellowship. A. Weltman gratefully acknowledges support from the South African Research Chairs Initiative of the Department of Science and Technology and the National Research Foundation of South Africa. J. Gordin, E. Platts and S. Kandhai are all supported by grantholder bursaries under this programme.  J. Gordin is financially supported by the National Astrophysics $\&$ Space Sciences Programme (NASSP). Any opinion, finding and conclusion or recommendation expressed in this material is that of the authors and the NRF does not accept any liability in this regard.
%  \newpage

\section{Tabulated Summary}
\label{TableSummary}
A living version of the following tabular summary can be found at \url{http://frbtheorycat.org}.

\begingroup
\setlength{\LTleft}{-20cm plus -1fill}
\setlength{\LTright}{\LTleft}
\begin{longtable}{|l|l|l|l|l|l|l|} 

\hline
& \textsc{Progenitor} & \textsc{Mechanism} & \textsc{Emission} & \textsc{Counterparts}  & \textsc{Type} & \textsc{References} 
\\\hline\hline

%%%%
\multirow{14}{*}{\rotatebox[origin=c]{90}{\textsc{Merger}}}
%%%%

& \multirow{3}{*}{NS--NS} & Mag. brak. & --- & GW, sGRB,  & Single  & \cite{Totani:2013lia}
\\\cline{3-4}\cline{6-7}
&  &  Mag. recon. & Curv. & afterglow, X-rays,  & Both  & \cite{Wang:2016dgs}
\\\cline{3-4}\cline{6-7}
&  &  Mag. flux & --- & kilonovae & Both & \cite{Dokuchaev:2017pkt}                             

\\\cline{2-7}
& NS--SN & Mag. recon. &  ---  &  None  &  Single  &  \cite{Egorov:2008yj} 

\\\cline{2-7} 
& \multirow{2}{*}{NS--WD} &  Mag. recon. &  Curv.  & --- &  Repeat  &  \cite{Gu:2016ygt}
\\\cline{3-7}
& & Mag. recon. & Curv. & --- & Single & \cite{Liu:2017ctq}

\\\cline{2-7}
& WD--WD  & Mag. recon. &  Curv.  &  X-rays, SN &  Single  &  \cite{Kashiyama:2013gza}

\\\cline{2-7}
& WD--BH & Maser &  Synch.  &  X-rays &  Single  &  \cite{Li:2018bjq}

\\\cline{2-7}
& NS--BH & BH battery & --- &  GWs, X-rays,  &  Single  &  \cite{Mingarelli:2015bpo} \\
& & & & $\gamma$-rays & & 
\\\cline{2-7}

& Pulsar--BH & --- & --- &  GWs &  Single  &  \cite{Bhattacharyya:2017orq}
\\\cline{2-7}

& KNBH--BH  & Mag. flux &  Curv.  & GWs, sGRB, & Single  &  \cite{Zhang:2016rli} \\
& (Inspiral) & & & radio afterglow& &

\\\cline{2-7}
&KNBH--BH & Mag. recon. &  Curv. & GW, $\gamma$-rays, & Single & \cite{Liu:2016olx} \\
& (Magneto.) & & & afterglow & &
\\\hline\hline

%%%%
\multirow{5}{*}{\rotatebox[origin=c]{90}{\textsc{Collapse}}}
%%%%
 
& NS to KNBH & Mag. recon. & Curv. & GW, X-ray & Single  & \cite{Falcke:2013xpa} \\
& & & & afterglow \& GRB & & \cite{Punsly:2016abn} \\
& & & & & & \cite{Zhang:2013lta}
\\\cline{2-7}

& NS to SS & $\beta$-decay & Synch. &  GW, X- \& $\gamma$-ray & Single & \cite{Shand:2015uda}
\\\cline{2-7}
& NS to BH & Mag. recon. &  Curv. & GW & Single & \cite{Fuller:2014rza}
\\\cline{2-7}

& SS Crust  & Mag. recon. &  Curv. & GW & Single & \cite{Zhang:2018zdn}
\\\hline\hline

%%%
\multirow{9}{*}{\rotatebox[origin=c]{90}{\textsc{SNR} (Pulsar)}}
%%%

& Giant Pulses & Various & Synch./  & --- & Repeat & \cite{Keane:2012yh} \\
 & & & Curv. &  & & \cite{Cordes:2015fua} \\
 & & & &  & & \cite{Connor:2015era}
\\\cline{2-7}
& Schwinger Pairs & Schwinger & Curv. & --- & Single & \cite{Lieu:2016hfw}
\\\cline{2-7}
& PWN Shock & --- & Synch. & SN, PWN,  & Single & 
\cite{Murase:2016sqo} \\
 & (NS) & & & X-rays & &
 \\\cline{2-7}
& PWN Shock & --- & Synch. & SN, X-rays & Single & \cite{Murase:2016sqo} \\
 & (MWD) & & & & &\\
\hline
%\hline
%%%
\multirow{6}{*}{\rotatebox[origin=c]{90}{\textsc{SNR} (Mag.)}}
%%%

& MWN Shock  & Maser & Synch. & GW, sGRB, radio & Single  & \cite{Popov:2007uv} \\
& (Single) & & & afterglow, high & & \cite{Murase:2016sqo} \\
& & & & energy  $\gamma$-rays & & \cite{Lyubarsky:2014jta} \\
\cline{2-7}
& MWN Shock & Maser& Synch. &  GW, GRB, radio & Repeat & \cite{Beloborodov:2017juh} \\
& (Clustered) & & & afterglow, high &  & \\
& & & & energy $\gamma$-rays &  &\\
\hline
\hline
 
\multirow{9}{*}{\rotatebox[origin=c]{90}{\textsc{AGN}}}

& Jet--Caviton & $e^-$ scatter & Bremsst. & X-rays, GRB, & Repeat  & \cite{Romero:2015nec}
\\\cline{6-7}
& & & & radio & Single & \cite{Vieyro:2017flr}
\\\cline{2-7}
& AGN--KNBH & Maser & Synch. & SN, GW, $\gamma$-rays, & Repeat & \cite{DasGupta:2017uac} \\
& & & & neutrinos & &
\\\cline{2-7}
& AGN--SS & $e^-$ oscill. & --- & Persistent GWs, & Repeat & \cite{DasGupta:2017uac}\\
& & & & GW, thermal rad., & & \\
& & & &  $\gamma$-rays, neutrinos& &
\\\cline{2-7}
& Wandering & --- & Synch. & AGN emission, & Repeat & \cite{Katz:2017ube} \\
& Beam & & & X-ray/UV & &\\
\hline\hline

\multirow{16}{*}{\rotatebox[origin=c]{90}{\textsc{Collision/Interaction}}}

& NS \& Ast./ &  Mag. recon. & Curv. & None & Single  &
\cite{Geng:2015vza} \\
& Comets& & & & & \cite{Huang:2015peq}
\\\cline{2-7}

& NS \& Ast. & $e^-$ stripping & Curv. &  $\gamma$-rays & Repeat  & \cite{Dai:2016qem} \\
& Belt & & & & & \cite{Bagchi:2017tzi}
\\\cline{2-7}

& Small Body  & Maser & Synch. & None & Repeat  & \cite{Mottez:2014awa} \\
& \& Pulsar & & & & & 
\\\cline{2-7}

& NS \& PBH & Mag. recon. & ---  & GW & Both  & \cite{Abramowicz:2017zbp}
\\\cline{2-7}

& Axion Star & $e^-$ oscill. & --- & None & Single & \cite{Iwazaki:2014wta,Iwazaki:2014wka,Iwazaki:2015zpb} \\
& \& NS & & & & & \cite{Raby:2016deh}
\\\cline{2-7}

& Axion Star & $e^-$ oscill. & --- & None & Repeat & \cite{Iwazaki:2017rtb} \\
& \& BH & & & & &
\\\cline{2-7}

& Axion Cluster  & Maser & Synch. & --- & Single & \cite{Tkachev:2014dpa} \\
& \& NS & & & & &
\\\cline{2-7}

& Axion Cloud  & Laser & Synch. & GWs & Repeat & \cite{Rosa:2017ury} \\
& \& BH & & & & &
\\\cline{2-7}

& AQN \& NS & Mag. recon. & Curv. & Below IR & Repeat  & \cite{vanWaerbeke:2018nyj}
\\\hline\hline

%%%
\multirow{17}{*}{\rotatebox[origin=c]{90}{\textsc{Other}}}
%%%

& Starquakes & Mag. recon. & Curv. & GRB, X-rays & Repeat  & \cite{Wang:2017agh}
\\\cline{2-7}

& Variable & Undulator & Synch. & --- & Repeat  & \cite{Song:2017yel} \\
& Stars & & & & &
\\\cline{2-7}

& Pulsar & Electrostatic & Curv. & --- & Repeat  & \cite{Katz:2017tcd} \\
& Lightning & & & & &
\\\cline{2-7}

& Wandering & --- & --- & --- & Repeat  & \cite{Katz:2016zxi} \\
& Beam & & & & &
\\\cline{2-7}

& Tiny EM  & Thin shell & Curv. & Higher freq.  & Repeat  & \cite{Thompson:2017wbo,Thompson:2017inw} \\
& Explosions & related &  & radio pulse, $\gamma$-rays & &
\\\cline{2-7}

& WHs & --- & --- & IR emission, $\gamma$-rays & Single  & \cite{Barrau:2014yka,Barrau:2018kyv}
\\\cline{2-7}

& NS Combing & Mag. recon. & --- & Scenario & Both  & \cite{Zhang:2017zse,Zhang:2018ueg}
\\\cline{2-7}

& Neutral Cosmic  & Cusp decay & --- & GW, neutrinos, & Single  & \cite{Brandenberger:2017uwo} \\
& Strings & & & cosmic rays, GRBs & &
\\\cline{2-7}

& Superconducting & Cusp decay & --- & GW, neutrinos, & Single  & \cite{Costa:2018zwb} \\
& Cosmic Strings & & & cosmic rays, GRBs & & 
\\\cline{2-7}

& Galaxy DSR & DSR & Synch. & ---  &  Both  & \cite{Houde:2017pni}
\\\cline{2-7}

& Alien Light& Artificial & --- & --- & Repeat  & \cite{Lingam:2017fwa} \\
& Sails & transmitter& & & &

\\\hline\hline

\multirow{4}{*}{\rotatebox[origin=c]{90}{\textsc{Inviable}}}

& Stellar Coronae & N/A & N/A  & N/A  & N/A & \cite{Loeb:2013wna} \\
& & & & & & \cite{Maoz:2015xpa}
\\\cline{2-7}

& Annihilating & N/A & N/A  & N/A  & N/A & \cite{Keane:2012yh} \\
& Mini BHs & & & & &
\\\hline
\caption{Tabulated Summary}
\label{Table}
\end{longtable}
\endgroup

\newpage
%\printglossaries
\printnoidxglossary[type=\acronymtype,numberedsection=autolabel]
\label{acronyms}

\newpage
\section{References}
\bibliographystyle{elsarticle-harv}
\bibliography{frb_bib}

\end{document}